\documentclass[twocolumn, aps, prd]{revtex4-2}
\usepackage{graphicx}% Include figure files
\usepackage{dcolumn}% Align table columns on decimal point
\usepackage{bm}% bold math
\usepackage{caption}
\usepackage{subcaption} %for subfigures

\usepackage[colorlinks=true, allcolors=black]{hyperref}
\usepackage[toc,page]{appendix}
\begin{document}

\preprint{APS/123-QED}

%working on title
\title{Stability of Thin Shell and Wormhole Configurations: Schwarzschild, Schwarzschild - (Anti-) de Sitter, and FLRW Spacetimes}% Force line breaks with \\

\author{Travis Seth Rippentrop}
\email{Travis.Rippentrop@utdallas.edu}
\author{Avijit Bera}
\email{avijit.bera@utdallas.edu}
\author{Mustapha Ishak}
\email{mishak@utdallas.edu}
\affiliation{Department of Physics, The University of Texas at Dallas, Richardson, TX 75080, USA}%Lines break automatically or can be forced with \\
%\author{Third Author}%
%\affiliation{%


%\affiliation{
%\affiliation{
%\affiliation{%


\date{\today}% It is always \today, today,

\begin{abstract}
The stability of thin shell wormholes and black holes to linearized spherically symmetric perturbations about a static equilibrium is analyzed. Thin shell formalism is explored and junctions formed from combinations of Schwarzschild, Schwarzschild - de Sitter, and Schwarzschild - anti-de Sitter, as well as Friedmann-Lemaître-Robertson-Walker (FLRW) spacetimes are considered. The regions of stability for these different combinations are thoroughly described and plotted as a function of mass ratios of the Schwarzschild masses and radii of the wormhole throats. A taxonomy of the qualitative features of the various configurations and parameter spaces is developed, illustrating the stability regions when present. The considered wormholes are all found to be unstable in the causal region.
% (\textit{i.e.}, $0<P'/\sigma'<1$)
%Write the abstract here. 
%\begin{description}


%\end{description}
\end{abstract}

%\keywords{Wormhole, Thin Shell, Stability Analysis, FLRW Schwarzschild Junction}%Use showkeys class option if keyword 
% display desired

\maketitle

% \tableofcontents

\section{Introduction \label{sec:Intro}}
\par\noindent

%The intro will be written later. 

Einstein's field equations allow for the mathematical existence of wormholes as exact solutions. A theoretical framework of constructing wormhole solutions is the thin shell or the Darmois-Israel formalism established in \cite{Darmois1927, Israel:1966rt} and used extensively elsewhere \cite{poisson1995thin, musgrave1996junctions, principles, ishak2002stability}. This method involves the assumption that the throat of a wormhole is infinitesimally short and the energy density therein is confined to an infinitesimally thin region known as the thin shell. Using these assumptions, it is possible to derive an equation of motion for the radius of the wormhole throat using the difference in the extrinsic curvature at the throat. From this, stability conditions can be derived using an effective potential based on the equation of motion of the throat, see e.g.  \cite{musgrave1996junctions,ishak2002stability}.
%Such explorations of stability are crucial in wormhole studies.

Detailed information on the thin-shell formalism in general relativity can be found in, e.g. \cite{principles, Lobo:2005zu, lobo2004shellstraversablewormholes} and references therein.
The thin shell approach has had wide applications in general relativity and has been the subject of many studies. An initial study of the stability of the thin shell Schwarzschild wormhole about a static solution can be found in \cite{poisson1995thin}. Since then many others have utilized similar techniques with varying spacetime metrics and conditions. Some studies have introduced a cosmological constant through the Schwarzschild - de Sitter, and Schwarzschild - anti-de Sitter spacetimes, e.g. \cite{ishak2002stability,Wang:2017gdp, Lemos:2004vs, Lobo:2003xd, Lemos:2003jb, salti2023thin}. Other studies have utilized charged wormholes (Reissner–Nordström) \cite{Eiroa:2008ky, Mazharimousavi:2015sga, Eiroa:2011nd, Eiroa:2015hrt, Kim_2001, Rahaman:2006vg}, rotating wormholes \cite{Tsukamoto:2018lsg, Kashargin:2011fg}, or Bardeen de-sitter wormholes \cite{Alshal_2024}. Furthermore, the stability of wormholes has also been considered for modified theories of gravity such as $F(R)$ \cite{Eiroa:2015hrt, Figueroa-Aguirre:2022qzf}, Einstein-Guass-Bonnet \cite{Kokubu:2015spa, Kokubu:2020lxs, Thibeault:2005ha}, and Hadara (Conformal Killing)  gravity \cite{Alshal_2024}. There has also been a study on the stability of various wormhole types when constrained by current cosmological observations \cite{Wang:2017gdp}. Surprisingly, fewer studies \cite{Sakai_1994, LA_CAMERA_2011, perez2023new, sahu2024singularhypersurfacesshellscosmology} have utilized the Friedmann-Lemaitre-Robertson-Walker metric, which we include in our paper along with other spacetimes. 

In this paper, we explore and systematize the study of the stability conditions of spherically symmetric thin shell spacetime junctions considering linearized perturbations about a static equilibrium. It is demonstrated that stability exists in the casual region (where the perturbation sound speed is real and sub-luminal) for black holes. However, for all wormhole constructions that are explored, stability is found to be present only far outside this region.

The structure of the paper is described as follows. In Section \ref{sec:Formalism}, we begin by establishing and summarizing the thin shell formalism, using the equation of motion of the radius of the throat to derive stability conditions.  In Section \ref{sec:Schwarzschild}, the stability conditions are applied to wormhole and black hole junctions composed of Schwarzschild, Schwarzschild - de Sitter, and Schwarzschild anti-de Sitter spacetimes. A taxonomy of major categories is defined, and mathematical conditions related to the geometry of the stability regions for each category are derived (this portion builds and expands on \cite{ishak2002stability}). Next, in Section \ref{sec: FLRW} we consider the construction of wormholes and black holes using the Friedmann–Lemaître–Robertson–Walker (FLRW) metric and a Schwarzschild or Schwarzschild - (anti-) de Sitter metric. We derive the extrinsic curvature and use the thin shell formalism to give and categorize the stability conditions of these junctions. The taxonomic conditions from the earlier section are generalized to apply to this latter section. In Section \ref{sec:results}, we present our analysis with two-dimensional plots and discussion. Finally, in Section \ref{sec:Conclusion}, a summary and concluding remarks are provided. Additionally, we display the stability regions in three-dimensional parameter space for a few configurations of both black holes and wormholes in Appendix \ref{sec:3d plots}.

\section{Formalism\label{sec:Formalism}}
\par\noindent

\begin{figure*}[hbt!]
     \centering
     \begin{subfigure}{0.95\textwidth}
         \centering
         % \captionsetup{labelformat=empty} 
         \includegraphics[width=\textwidth]{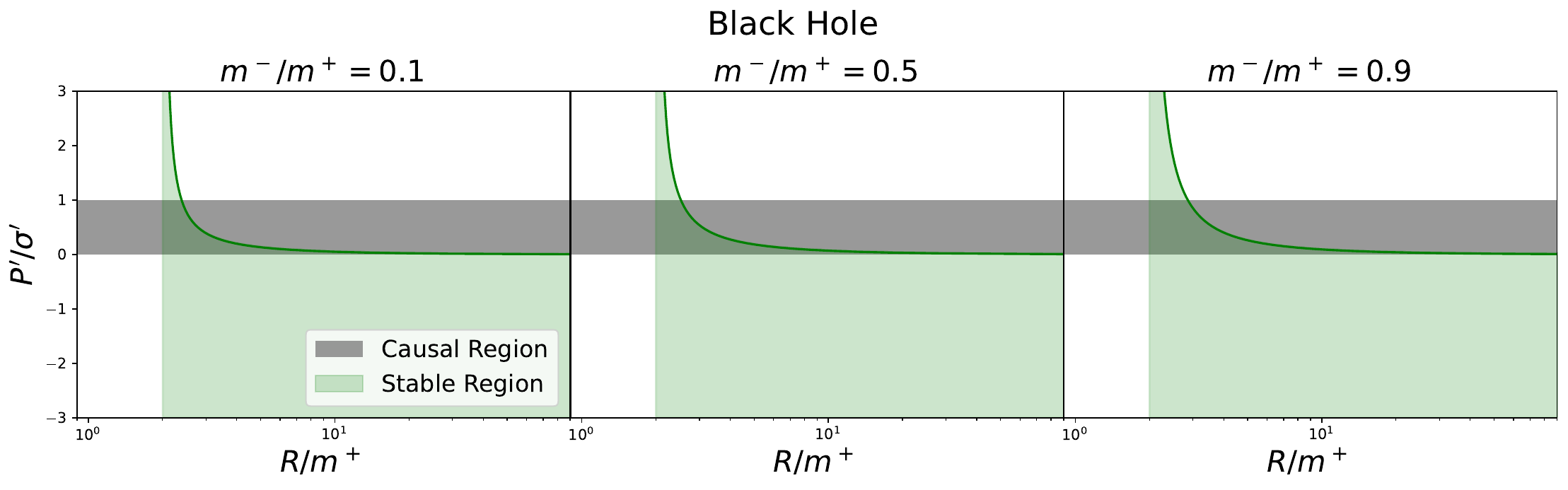}
         % \caption{} 
         % \label{fig:SS_BH}
     \end{subfigure}
     \begin{subfigure}{0.95\textwidth}
         \centering
         % \captionsetup{labelformat=empty} 
         \includegraphics[width=\textwidth]{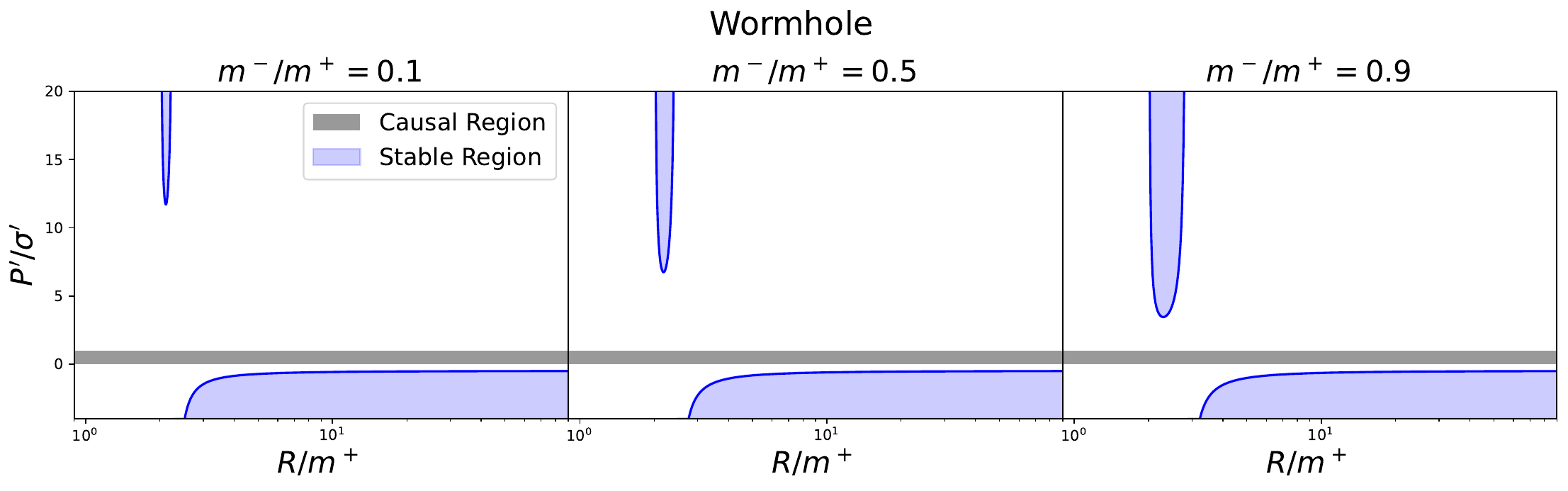}
         % \caption{} 
         % \label{fig:SS_WH}
     \end{subfigure}
     \hfill
        \caption{\textbf{Schwarzschild -- Schwarzschild Junction:} The areas shaded in green and blue correspond to the stability regions for black hole and wormhole, respectively. The shaded region in gray represents the causal region in both cases. This color scheme is used consistently throughout this paper. For the black hole, $M\neq0$ anywhere and there is no stability flip. Stability partially intersects the Causal Region. For the wormhole there is an asymptote present and stability is separated into two disjoint regions, neither of which intersect the Causal Region. For 3-D plot, see figure (\ref{fig:SS_3d}).}
        \label{fig:SS}
\end{figure*}

\begin{figure*}[t]
     \centering
     \begin{subfigure}[b]{0.95\textwidth}
         \centering
         \includegraphics[width=\textwidth]{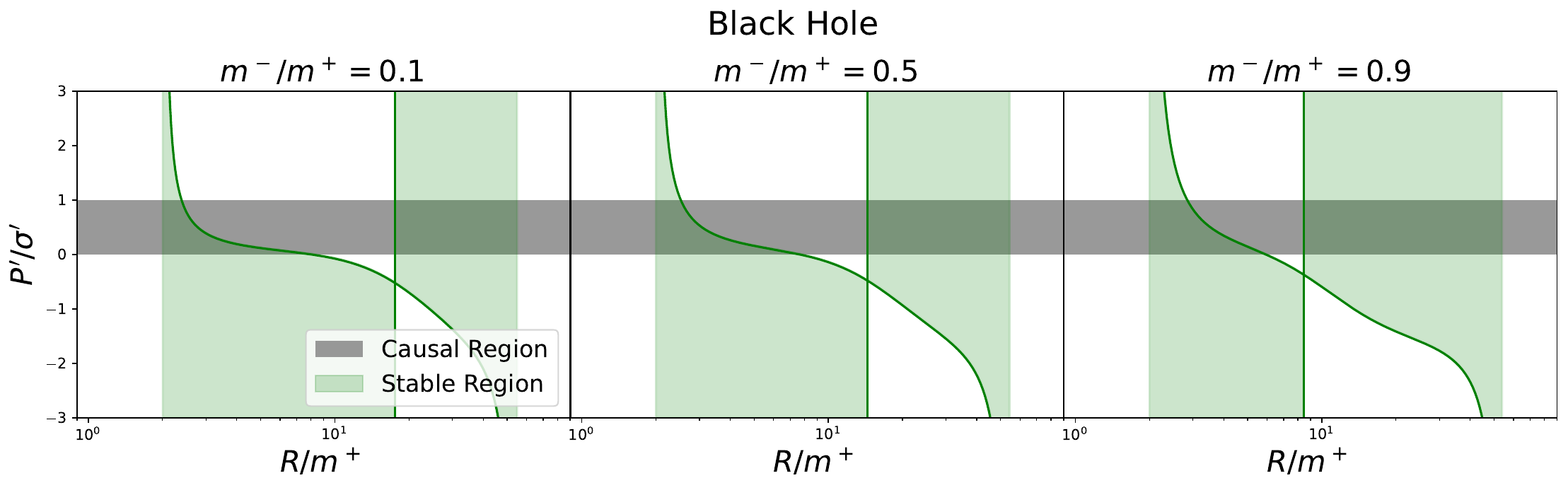}
         %\label{fig:BH}
     \end{subfigure}
     \begin{subfigure}[b]{0.95\textwidth}
         \centering
         \includegraphics[width=\textwidth]{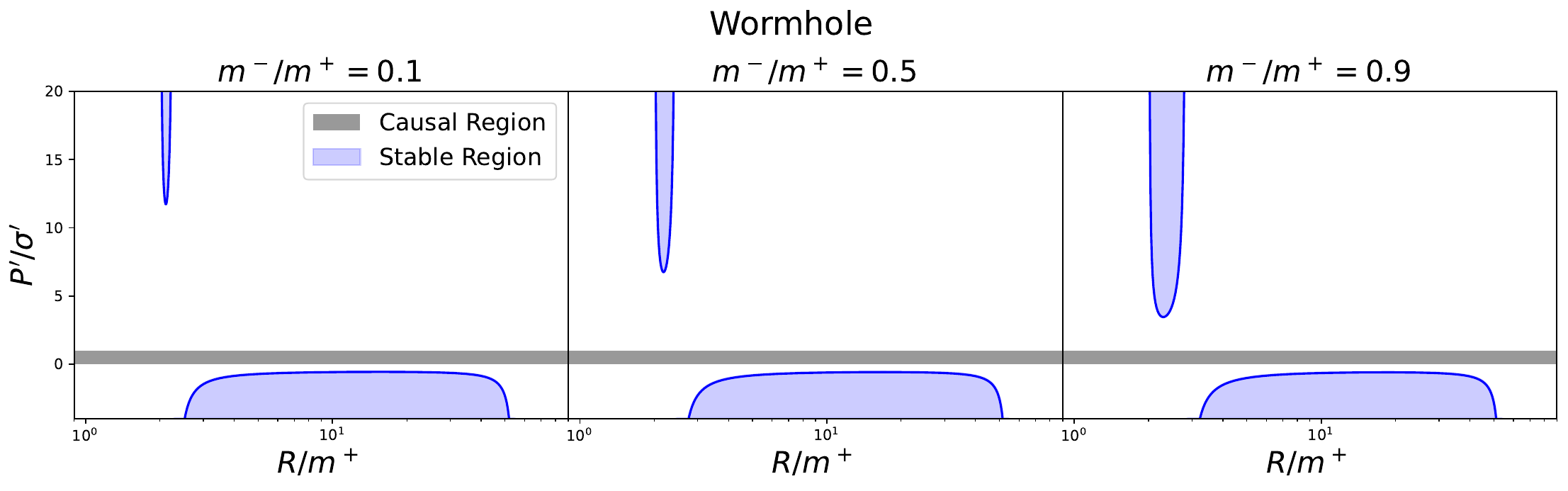}
         %\label{fig:BH}
     \end{subfigure}
     \hfill
    \caption{\textbf{Schwarzschild - de Sitter -- Schwarzschild Junction:} For the black hole, there are points where $M=0$ and a stability flip is present without an asymptote. This occurs at the vertical line. A Schwarzschild - de Sitter -- Schwarzschild junction implies $[C] =-0.001$ which does not fulfill the asymptote condition for any beta. For both plots the effects of the de Sitter Horizon can be seen by the divergence around $\alpha \approx 54$. For 3-D plot, see figure (\ref{fig:SdsS_3d})}
    \label{fig:SdsS}
\end{figure*}

\begin{figure*}[t]
     \centering
     \begin{subfigure}[b]{0.95\textwidth}
         \centering
         \includegraphics[width=\textwidth]{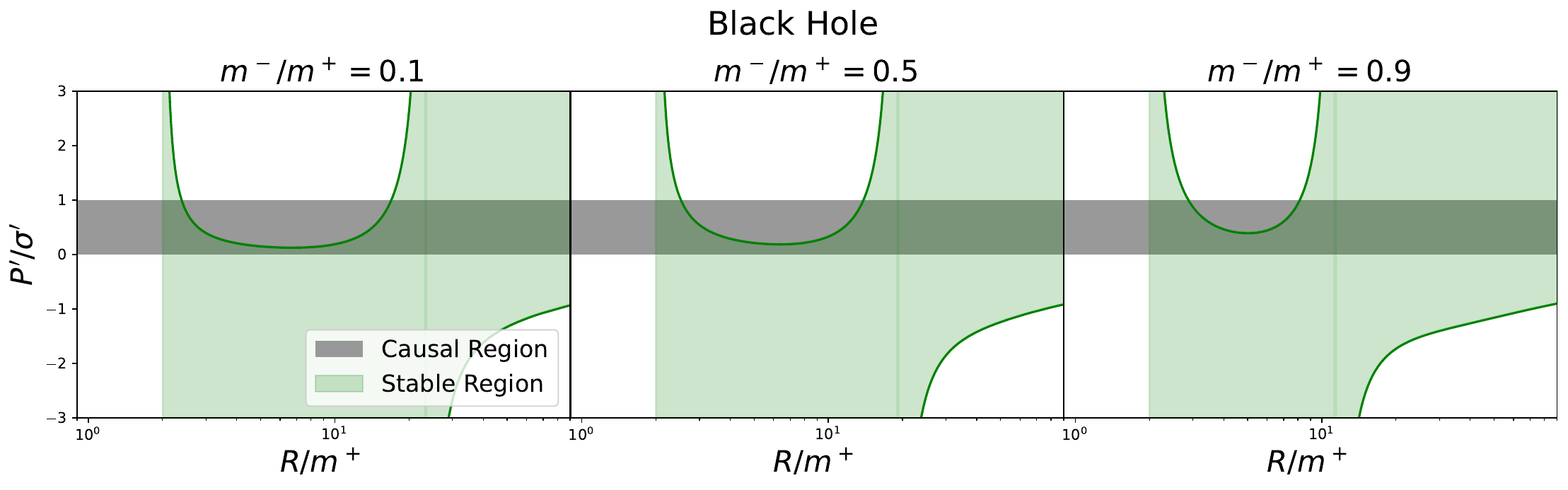}
         %\label{fig:BH}
     \end{subfigure}
     \begin{subfigure}[b]{0.95\textwidth}
         \centering
         \includegraphics[width=\textwidth]{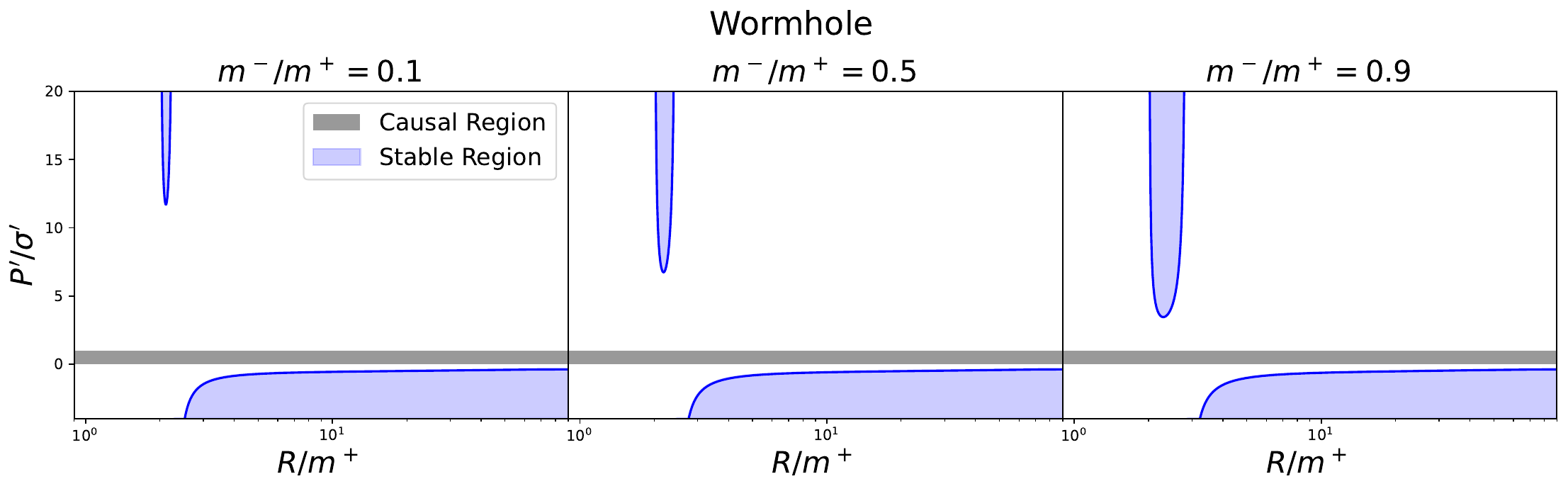}
         %\label{fig:WH}
     \end{subfigure}
     \hfill
        \caption{\textbf{Schwarzschild Anti-de Sitter -- Schwarzschild Junction:} Black hole has an asymptote as $[C] =0.001$ which does fulfill the asymptote condition. A de Sitter horizon is not present as $\Lambda^\pm\leq0$. For 3-D plot, see figure (\ref{fig:SadsS_3d}).}
        \label{fig:SadsS}
\end{figure*}

\begin{figure*}[t]
     \centering
     \begin{subfigure}[b]{0.95\textwidth}
         \centering
         \includegraphics[width=\textwidth]{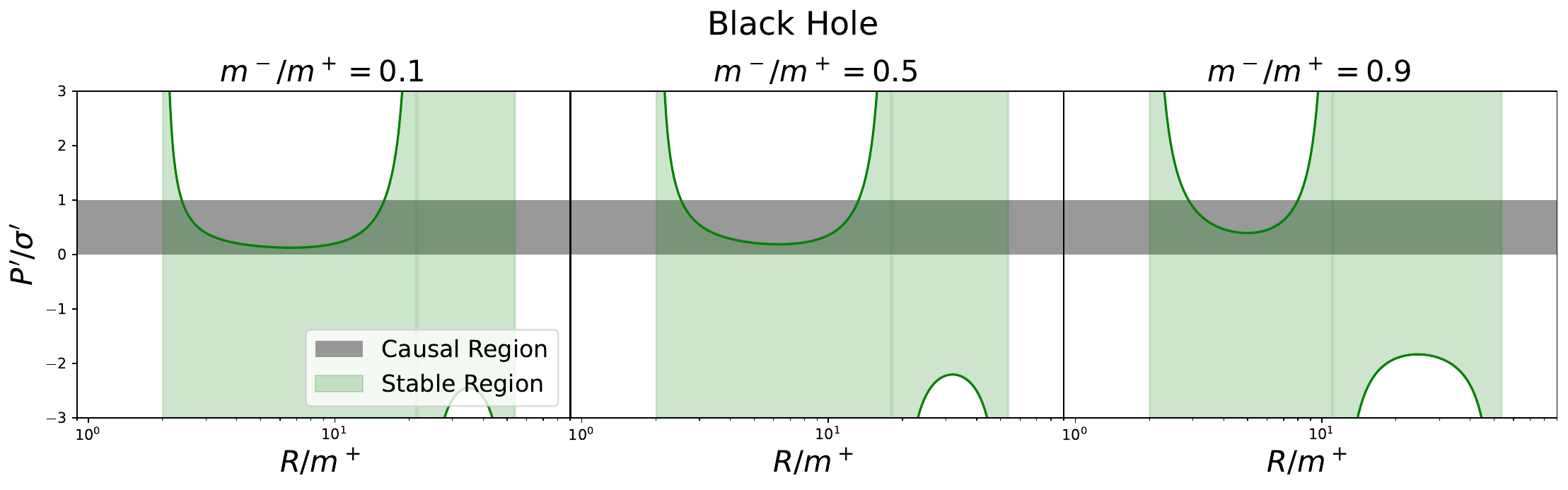}
         %\label{fig:BH}
     \end{subfigure}
     \begin{subfigure}[b]{0.95\textwidth}
         \centering
         \includegraphics[width=\textwidth]{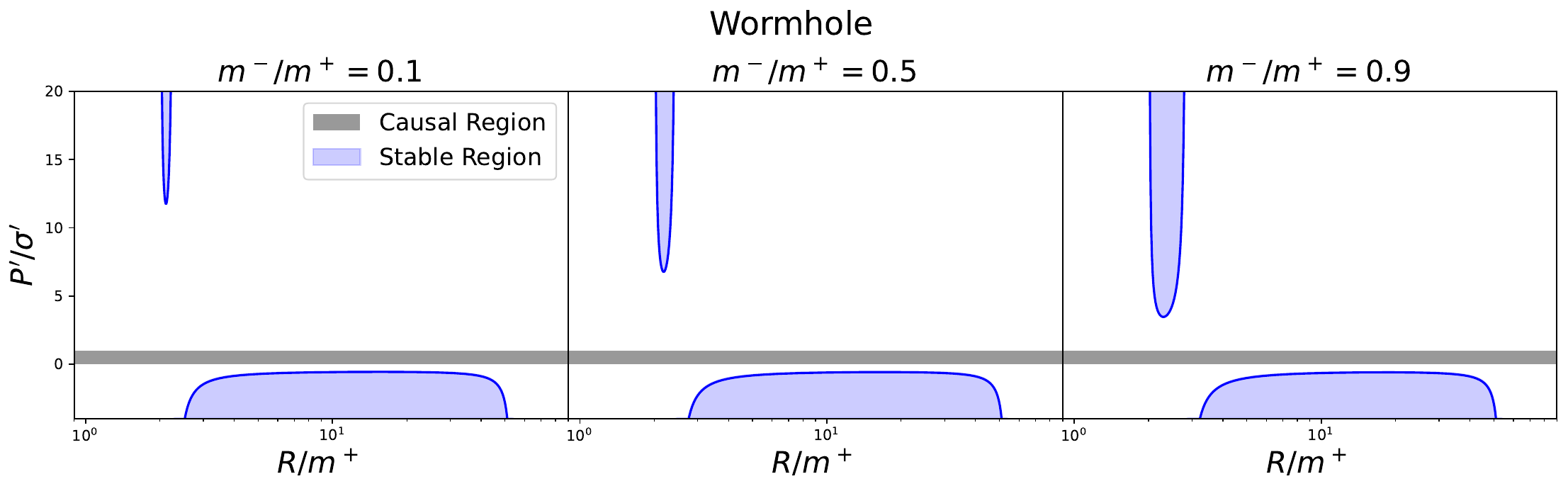}
         %\label{fig:WH}
     \end{subfigure}
     \hfill
        \caption{\textbf{Schwarzschild -- Schwarzschild - de Sitter  Junction:} Black hole is similar to Schwarzschild - anti-de Sitter -- Schwarzschild (figure \ref{fig:SadsS}) with a similar asymptote, though in this case the de Sitter Horizon is also present. Again $[C] =0.001$ in this case, fulfilling the asymptote condition. De Sitter Horizon is present.}
        \label{fig:SSds}
\end{figure*}

\begin{figure*}[t]
     \centering
     \begin{subfigure}[b]{0.95\textwidth}
         \centering
         \includegraphics[width=\textwidth]{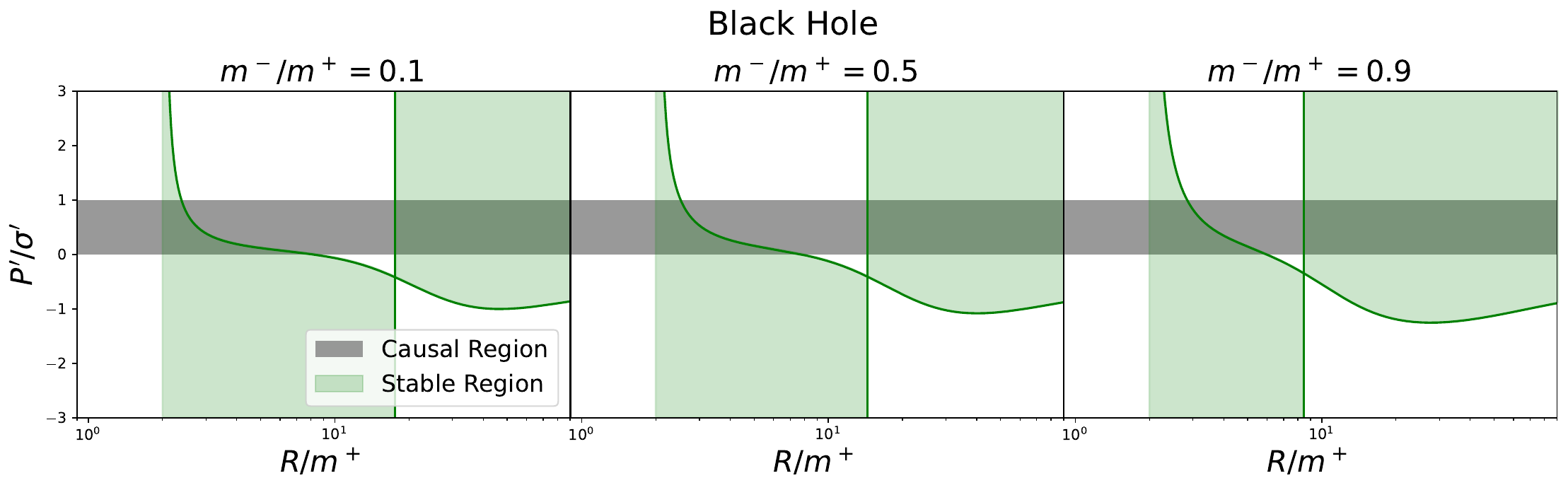}
         %\label{fig:BH}
     \end{subfigure}
     \begin{subfigure}[b]{0.95\textwidth}
         \centering
         \includegraphics[width=\textwidth]{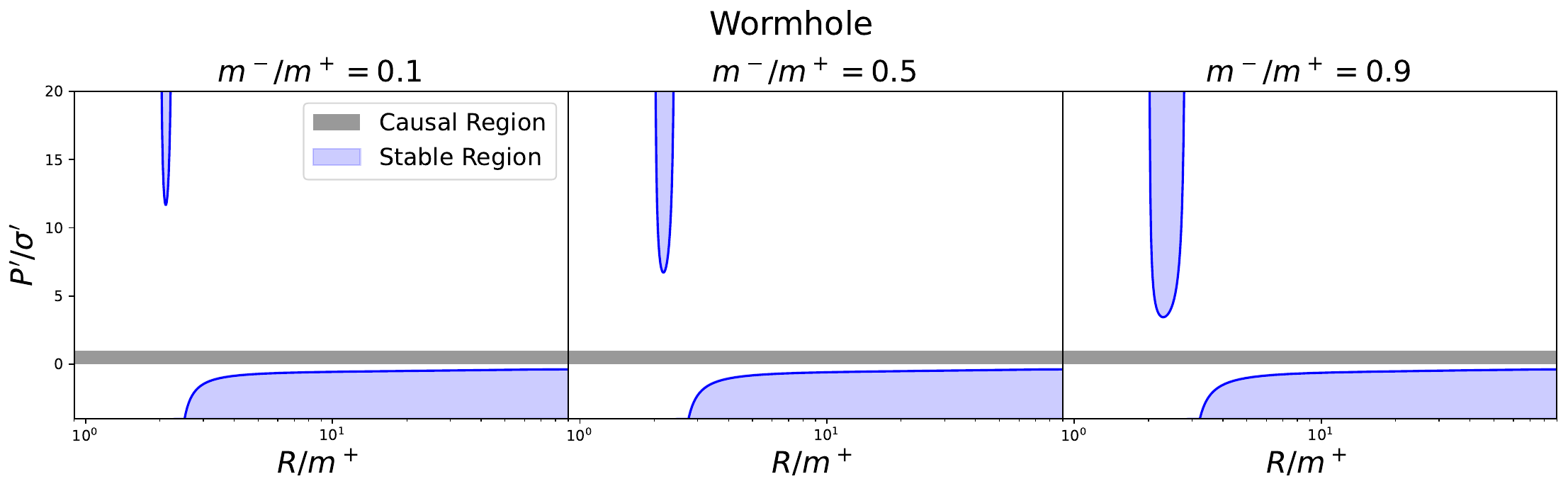}
         %\label{fig:WH}
     \end{subfigure}
     \hfill
        \caption{\textbf{Schwarzschild -- Schwarzschild - anti-de Sitter Junction:} Black hole is similar to Schwarzschild - de Sitter -- Schwarzschild (figure \ref{fig:SdsS}) but does not have a de Sitter Horizon. $[C]=-0.001$ and there is no asymptote.}
        \label{fig:SSads}
\end{figure*}

% \ref{fig:SS_BH}
We consider two spacetimes $\mathcal{M}^+$ and $\mathcal{M}^-$ defined by metrics $g_{\alpha\beta}^+$ and $g_{\alpha\beta}^-$. We define two hypersurfaces within each spacetime as $\Sigma^+$ and $\Sigma^-$ with intrinsic metrics of $g_{ij}^+$ and $g_{ij}^-$, respectively. $x_\pm^\gamma$ refers to the coordinates in $g_{\alpha\beta}^\pm$ and $\xi_\pm^c$ refers to the coordinates in $g_{ij}^\pm$.
The parametric equation of the surface takes the form $F(x^\alpha(\xi^a))=0$ \cite{musgrave1996junctions}.

Throughout this work, we define $[A] \equiv A^+-A^-$ and $\bar{A} \equiv \frac12 (A^++A^-)$ for some quantity $A$.
The first Darmois condition for joining a portion of $\mathcal{M}^+$ to a portion of $\mathcal{M}^-$ is (see, e.g., \cite{musgrave1996junctions})
\begin{equation}
[g_{ij}]=0.
\label{1st darmois}
\end{equation}
This implies $g_{ij}^+=g_{ij}^-=g_{ij}$ and $\Sigma^+=\Sigma^-=\Sigma$.

% Coordinates $\xi^c$ will be defined as $(\tau,\theta,\phi)$ and $x^\gamma$ will be $(t,r,\theta,\phi)$. $\dot{X}$ will denote $\frac{dX}{d\tau}$.

% For most of this work we shall adopt $c=G=1$ units.

\begin{figure*}
     \centering
     \begin{subfigure}[b]{0.95\textwidth}
         \centering
         \includegraphics[width=\textwidth]{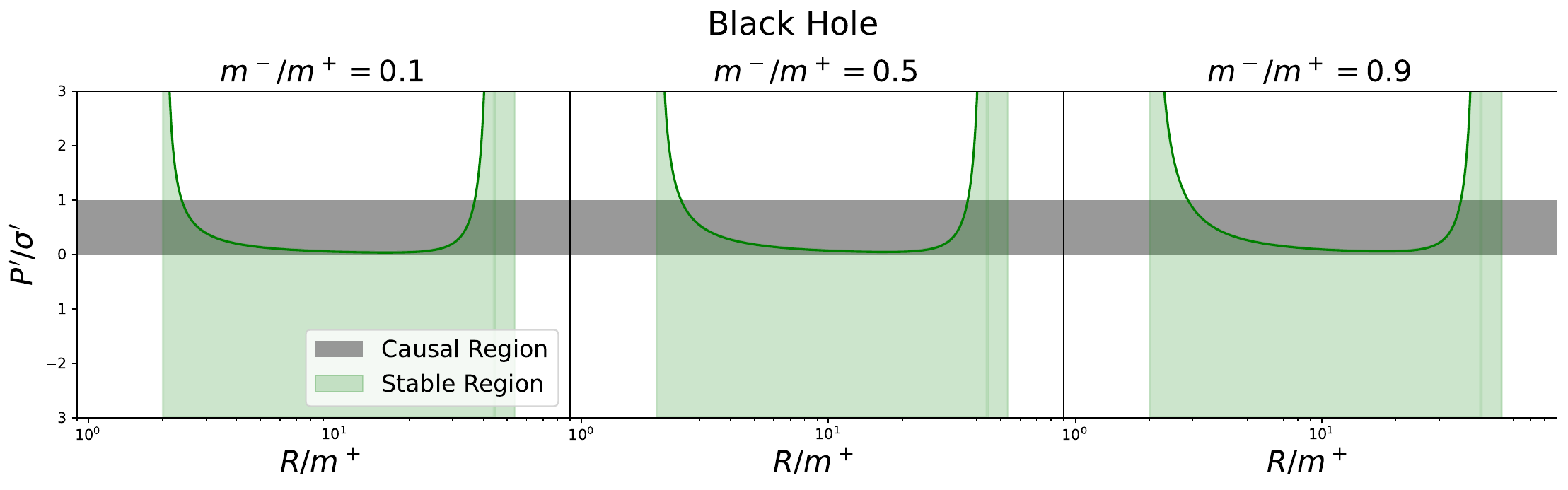}
         %\label{fig:BH}
     \end{subfigure}
     \begin{subfigure}[b]{0.95\textwidth}
         \centering
         \includegraphics[width=\textwidth]{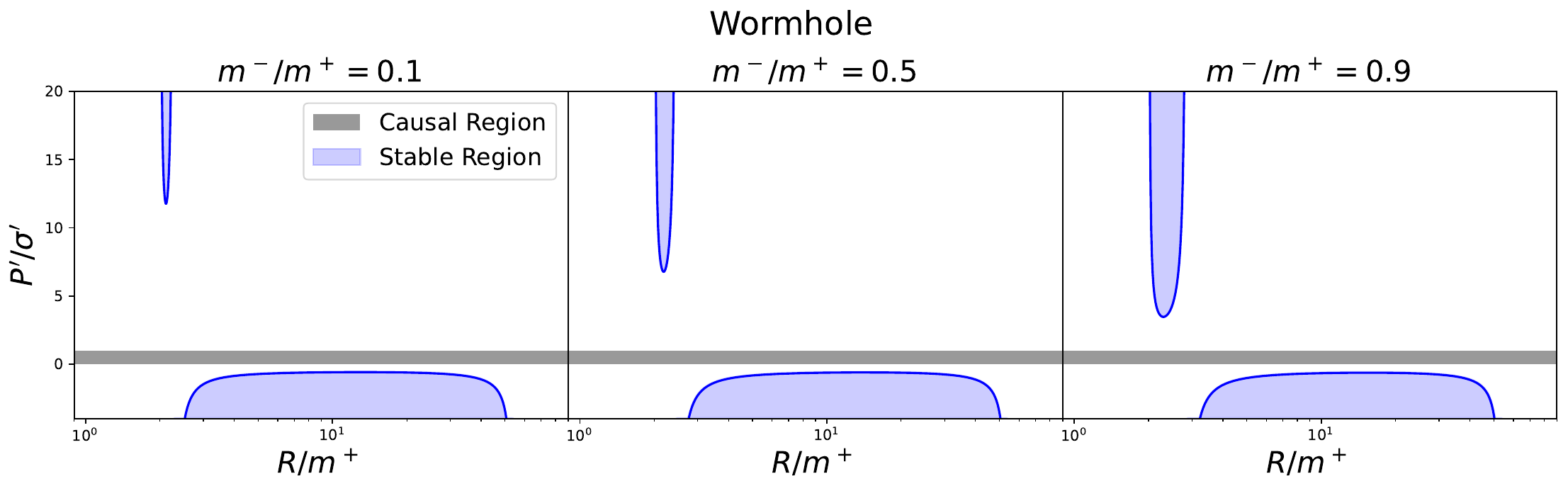}
         %\label{fig:WH}
     \end{subfigure}
     \hfill
        \caption{\textbf{Schwarzschild - de Sitter -- Schwarzschild - de Sitter Junction:} Black hole still possesses asymptote as $[C]=0$ fulfills asymptote condition in equation (\ref{asymptote condition}) if there exists a $\Lambda>0$. The location and shape of the asymptote differs from other cases and occurs at much higher $R$. Black hole plot has been expanded to view unstable region at $P'/\sigma'<-12$.}
        \label{fig:SdsSds}
\end{figure*}
The coordinates $(\tau,\theta,\phi)$ are indicated by $\xi^c$ and $(t,r,\theta,\phi)$ are indicated by $x^\gamma$. $\dot{A}$ is defined as the derivative of $A$ with respect to proper time $\tau$. We adopt $c=G=1$ units throughout this work.%(i.e., $\dot{X} \equiv \frac{dX}{d\tau})$. 

Per the thin shell approach, we let the throat of the wormhole be infinitesimally small and let each manifold have a boundary at the surface. In the case of a time-like spherically symmetric surface of dynamic radius $R(\tau)$ the surface line element can be written as \cite{ishak2002stability}
\begin{equation}
ds^2_\Sigma = -d\tau^2+R^2(\tau)d\Omega^2,
\end{equation}
where the surface is defined by $r=R(\tau)$ and is parameterized by the function $F(r)=r-R(\tau)=0$.

This boundary causes a discontinuity in the extrinsic curvature (second fundamental form) of the union of $\mathcal{M}^+$ and $\mathcal{M}^-$. The stress-energy tensor ($S_{ij}$) for this boundary can be calculated using the Lanczos equation, which is given by \cite{musgrave1996junctions}
\begin{equation}
S_{ij} = -\frac1{8\pi}([K_{ij}]-g_{ij}[K^i_i]),
\end{equation}
where $K_{ij}$ is the extrinsic curvature and is given by
\begin{equation}
    K_{ij}=-n_\gamma \bigg(\frac{\partial^2x^\gamma}{\partial\xi^i\partial\xi^j} + \Gamma^\gamma_{\alpha\beta} \frac{\partial x^\alpha}{\partial\xi^i} \frac{\partial x^\beta}{\partial\xi^j}\bigg).
\end{equation}
Note that $n_\gamma$ is the unit 4-normal to the surface $\Sigma$ in manifold $\mathcal{M}$ and is expressed as \cite{musgrave1996junctions}
\begin{equation}
    n_\gamma = \pm \frac{1}{\big( \big | g^{\alpha\beta}\frac{\partial F}{\partial x^\alpha} \frac{\partial F}{\partial x^\beta}\big |\big)^{1/2}} \frac{\partial F}{\partial x^\gamma},
\end{equation}
where the sign of $n_\gamma$ depends on the direction of the normal vector.

We treat $S_{ij}$ analogously to a perfect 4-fluid with $S_{ij}={\rm diag}(-\sigma,P,P)$ \cite{poisson1995thin}.
It can be shown that the energy density is
\begin{equation}
\sigma(\xi^a) = -S_\tau^\tau= -\frac{1}{4\pi}\big[K_\theta^\theta \big].
\end{equation}

We define the mass of the thin shell as
\begin{equation}
M=4\pi R^2\sigma=-\big[K_{\theta\theta} \big].
\label{M definition}
\end{equation}

If equation (\ref{1st darmois}) holds and $[K_{\theta\theta}] = 0$ then we refer to $\Sigma$ as a boundary surface, if $[K_{\theta\theta}] \neq 0$ then $\Sigma$ is a thin shell.

\footnotetext{The horizon occurs at $R/m^+ \approx 54$ which is close to $\alpha_{\rm dS}\approx54.77$ for $C=0.001$ -- see text}

In Section \ref{sec:Schwarzschild}, we consider a junction between two spacetimes of the form,
\begin{equation}
    ds_\pm^2=-\bigg(1-\frac{2\mu^\pm(r)}{r}\bigg)dt^2+\frac{dr^2}{1-\frac{2\mu^\pm(r)}{r}}+r^2d\Omega^2,
\label{standard metric form}
\end{equation}
where $\mu^\pm(r)$ represents the effective mass contained within the thin shell and is defined as \cite{musgrave1996junctions}
\begin{equation}
\mu^\pm(r)=\frac{1}{2} (g^\pm_{\theta\theta})^{\frac{3}{2}} R_{\theta\phi}^{\quad\theta\phi},
\label{effective mass}
\end{equation}
where metric tensor element $g_{\theta \theta}$ and Riemann tensor element $R_{\theta \phi}^{\quad\theta\phi}$ are computed on $\mathcal{M}$ (not on $\Sigma$).

Using equation (\ref{M definition}) we get
\begin{equation}
    M(R) = wR\sqrt{1-\frac{2\mu^-}{R}+\dot{R}} - R\sqrt{1-\frac{2\mu^+}{R}+\dot{R}},
\label{normal metric}
\end{equation}
where $w$ is determined by the direction of the normal vectors. If the vectors point in the opposite directions $w=-1$ and the junction is referred to as a wormhole. If the vectors point in the same direction $w=1$ and the junction is referred to as a black hole.

 Rearranging equation (\ref{normal metric}) gives the equation of motion \cite{musgrave1996junctions}
\begin{equation}
\dot{R}^2=\bigg(\frac{[\mu]}{M}\bigg)^2+\frac{2\bar{\mu}}{R}+\bigg(\frac{M}{2R}\bigg)^2-1.
\label{evolution eq}
\end{equation}

If $\mu^\pm(R)$ is defined uniquely for each $R$ we can define a potential $V(R)=-\dot{R}^2$ \cite{ishak2002stability}. 
We can expand this potential to second order about a static solution at $R_0=\rm const$. \cite{poisson1995thin}
\begin{equation}
V(R)\approx V(R_0)+V'(R_0)(R-R_0)+\frac12 V''(R_0)(R-R_0)^2.
\end{equation}

For this static solution, $V(R_0)=V'(R_0)=0$. A stable solution is given by $V''(R_0) > 0$. It follows from the definition of $V(R)$ that the equilibrium condition $V'(R)=0$ for a static solution is 
\begin{equation}
    \bigg(\frac{M}{2R} \bigg)' = -\frac{2R}{M}\bigg(\bigg(\frac{[\mu]}{R}\bigg)\bigg(\frac{[\mu]}{R}\bigg)'+\bigg(\frac{\bar{\mu}}{R}\bigg)'\bigg) \equiv \Gamma.
\end{equation}

The condition for a stable equilibrium ($V''(R)>0$) is 
\begin{equation}
\bigg(\frac{M}{2R}\bigg)\bigg(\frac{M}{2R}\bigg)'' < \Psi-\Gamma^2,
\label{stability equi}
\end{equation}
where $\Psi = -\big(\frac{[\mu]}{M}\big)'^2-\big(\frac{[\mu]}{M}\big)\big(\frac{[\mu]}{M}\big)''-\big(\frac{\bar{\mu}}{R}\big)'' $ \cite{ishak2002stability}.

In general, the conservation identity must also be satisfied and is given by
\begin{equation}
    \nabla_iS^i_j=- \bigg[T_\alpha^\beta\frac{\partial x^\alpha}{\partial \xi^j}n_\beta \bigg],
\end{equation}
which yields,
\begin{equation}
\dot{\sigma}=-2\frac{\dot{R}}{R}(\sigma+P)+ \Xi,
\end{equation}
where the flux term $\Xi$ is 
\begin{equation}
\Xi\equiv\bigg[T_\alpha^\beta\frac{\partial x^\alpha}{\partial \tau}n_\beta \bigg].
\end{equation}
For any vacuum solution, flux term $\Xi=0$, and the conservation identity becomes
\begin{equation}
\sigma'=-\frac{2}{R}(\sigma+P),
\end{equation}
which can be rewritten as,
\begin{equation}
    \bigg(\frac{M}{2R}\bigg)'' = \frac{\Upsilon}{2R^3}\bigg(1+2\frac{P'}{\sigma'}\bigg),
    \label{transparency}
\end{equation}
where $\Upsilon = 3M - (MR)'$.
Plugging this into the stability condition (equation \ref{stability equi}) yields
\cite{ishak2002stability}
\begin{eqnarray}
    \frac{P'}{\sigma'} &<& \frac12(\Phi-1); \quad M\Upsilon > 0, \rm and
    %\label{stability condition 1}
    \\ 
    \frac{P'}{\sigma'} &>& \frac12(\Phi-1); \quad M\Upsilon < 0,
    %\label{stability condition 2}
\end{eqnarray}
where $\Phi=\frac{4R^4}{M\Upsilon}(\Psi-\Gamma^2)$.

However, it is not necessary to compute $\Phi$. When using the condition $V'=0$ one can show that the stability conditions become
\begin{eqnarray}
    \frac{P'}{\sigma'} &<& \frac{R^3}{\Upsilon}\bigg(\frac{M}{2R}\bigg)''-\frac12; \quad M\Upsilon>0 \rm;
    \label{stability condition 1}
    \\ 
    \frac{P'}{\sigma'} &>& \frac{R^3}{\Upsilon}\bigg(\frac{M}{2R}\bigg)''-\frac12; \quad M\Upsilon < 0.
    \label{stability condition 2}
\end{eqnarray}

Note that the stability regions lie above and below the surface mapped out by the transparency condition (equation \ref{transparency}).

Stability within the region defined by $0 \leq \frac{P'}{\sigma'} < 1$ is of particular interest as it is the causal region. $P'/\sigma'$ corresponds to the square of a sound speed of perturbations. Thus, for physical solutions it seems natural to restrict this sound speed to real, sub-luminal values, where it is causal. The legitimacy of this assumption is discussed in the conclusion.

\section{Schwarzschild and (Anti-) de Sitter spacetimes}
\label{sec:Schwarzschild}
We begin our stability analysis by considering Schwarzschild wormhole and black hole solutions. We also consider the effects of a positive and negative cosmological constant corresponding to Schwarzschild - de Sitter and Schwarzschild - anti-de Sitter spacetime respectively.

A Schwarzschild spacetime is one where $\mu^\pm(R)=m^\pm$ where $m^\pm$ is the Schwarzschild mass.
A Schwarzschild - de Sitter spacetime is one where $\mu^\pm(R)=m^\pm+\frac{\Lambda^\pm}{6}R^3$ and $\Lambda^\pm>0$.
And a Schwarzschild - anti-de Sitter spacetime is like the above but with $\Lambda^\pm<0$.

Applying the condition of a static solution with constant $R$, we can set $V=-\dot{R}=0$.

We now have $M$ given by
\begin{equation}
    M = w\sqrt{1-\frac{2m^-}{R}-\frac{\Lambda^-}{3}R^2}-\sqrt{1-\frac{2m^+}{R}-\frac{\Lambda^+}{3}R^2}.
    \label{Sch M}
\end{equation}

From these three possibilities we get 18 junction combinations (9 wormholes and 9 black holes). A taxonomy of the qualitative properties of the stability conditions, (equations \ref{stability condition 1}, and \ref{stability condition 2}) can be created by looking at two main features.
The first is the existence of an asymptote at $\Upsilon=0$. The second is the limiting behavior as $R\rightarrow \infty$, whether or not the surface diverges at some finite value.
Finally, it is also worth noting the presence of a stability flip (the stability region shifting from above the surface to below or vice versa) caused by $M$ or $\Upsilon$ switching signs.

Throughout the rest of this paper where relevant, we only consider cases where $m^-<m^+$, which is reflected in our axis limits of our plots $0<m^-/m^+<1$  \cite{ishak2002stability} (in general $\mu^-$ and $\mu^+$ can be swapped without consequence to the stability regions).

%From the results, several patterns emerge.
%Firstly, for all combinations, there exists a region $R<2\mu_{\rm max}$ where $M$ and thus also $P'/\sigma'$ is complex. We refer to the boundary of this region as the ``low $R$ cutoff". Physically, this corresponds to a thin shell with a smaller radius than the largest Schwarzschild radius of the two spacetimes. In such a case, the thin shell would be within an event horizon.

First, it is important to note the region for which our analysis cannot yield results. When $R<2\mu_{\rm max}$ the radius of our thin shell is within the event horizon of one of the Schwarzschild space times and due to our use of Schwarzschild coordinates, we cannot analyze stability. Similarly, when at least one Schwarzschild - de Sitter spacetime is used, a cosmological horizon known as the de Sitter horizon is present at high $R$. This causes the surface to diverge at finite $R$. Since $R$ is large at this horizon we can estimate its value as $R_{\rm dS}\approx\sqrt{3/\Lambda_{\rm max}}$ where $\Lambda_{\rm max}$ is the greatest $\Lambda$ involved in the junction. Beyond the horizon, we reach another region where our stability analysis breaks down.
%Since $\Lambda$ can be thought of as a constant energy density, increasing the radius $R$ of the thin shell increases the amount of matter within that shell. Eventually the contained matter will be so great, its Schwarzschild radius would become larger than the radius of the thin shell and the thin shell would once again be inside an event horizon. This is exactly the limit this cutoff corresponds to.

An asymptote between these two horizons occurs if $\Upsilon=0$ \cite{ishak2002stability}. A wormhole with any combination of the above spacetimes always contains such an asymptote. A black hole may or may not contain an asymptote based on the following.

It can be seen from plotting that for small $R$ (near the event horizon) $\Upsilon$ and $M$ both have the same sign. With the exception of Schwarzschild -- Schwarzschild (figure \ref{fig:SS}), after a certain point (before $\Upsilon=0$ or $M=0$) if $\Upsilon$ monotonically increases, $M$ monotonically decreases or vice versa. Thus, only one of these quantities can be equal to zero for a certain junction. If $M=0$ anywhere then $\Upsilon \neq 0$ and there is no asymptote, though a stability flip would still exist at $M=0$. This lets the simpler condition $M=0$ become an indicator of the nonexistence of an asymptote.

$M=0$ implies $\mu^+=\mu^-$ or
\begin{equation}
\frac{6[m]}{R^3}=-[\Lambda].
\label{M=0 condition}
\end{equation}

Since $R$ must be positive, if $[m]$ and $[\Lambda]$ have the same sign, the condition can never be satisfied for any $R$, and so an asymptote must exist. This gives an asymptote existence condition of
\begin{equation}
\frac{[\Lambda]}{[m]}>0.
\label{Lam < 0 asymptote condition}
\end{equation}

For a junction with at least one $\Lambda>0$ there is an upper limit to $R$. If the $R$ value which satisfies the horizon condition (equation \ref{M=0 condition}) is greater than $R_{\rm dS}$ then the condition for asymptote non-existence cannot be met and an asymptote would be present. Taking $R_{\rm dS} \approx \sqrt{3/\Lambda_{\rm max}}$ gives an asymptote existence condition of
\begin{equation}
\frac{[\Lambda]}{[m]}>\frac{-2\Lambda_{\rm max}^\frac32}{\sqrt3}.
\label{asymptote condition}
\end{equation}

For both $\Lambda \leq 0$ this restriction does not exist and only equation (\ref{Lam < 0 asymptote condition}) applies.

For the Schwarzschild -- Schwarzschild case (figure \ref{fig:SS}), $M$ and $\Upsilon$ approach zero for large $R$, but never reach it. Thus, there is neither an asymptote or stability flip in this simple case.

It is also apparent from plotting that $M\Upsilon$ monotonically decreases for black holes and monotonically increases for wormholes. This behavior dictates the location of stability regions, whether above or below the surface.

\begin{figure*}
     \centering
     \begin{subfigure}[b]{0.95\textwidth}
         \centering
         \includegraphics[width=\textwidth]{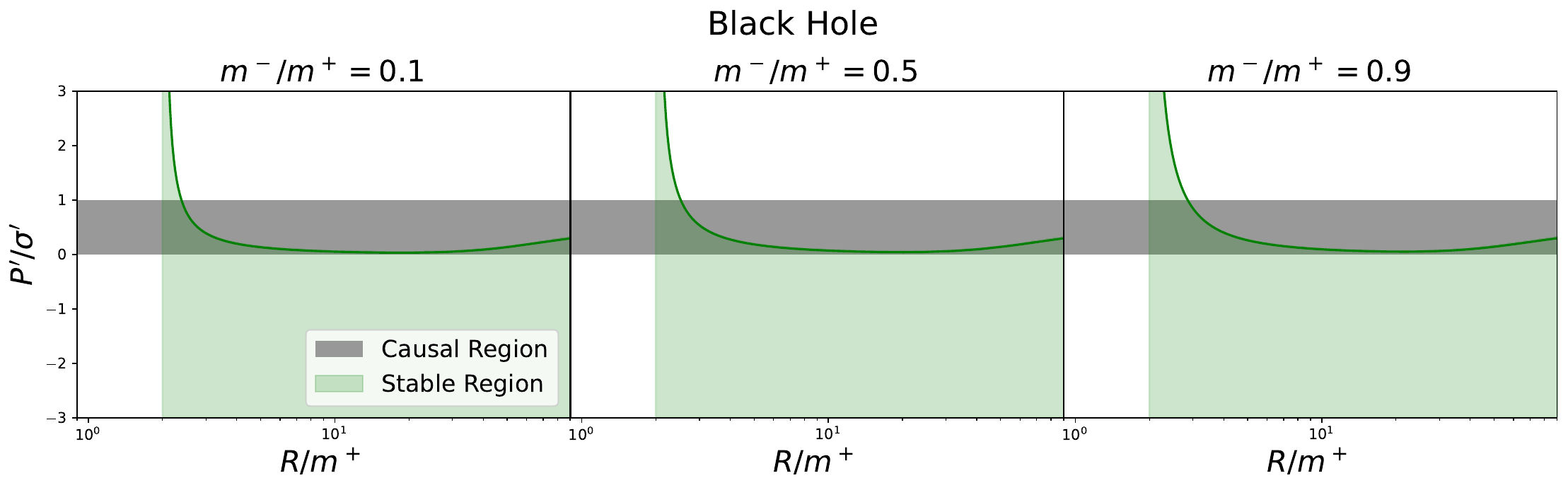}
         %\label{fig:BH}
     \end{subfigure}
     \begin{subfigure}[b]{0.95\textwidth}
         \centering
         \includegraphics[width=\textwidth]{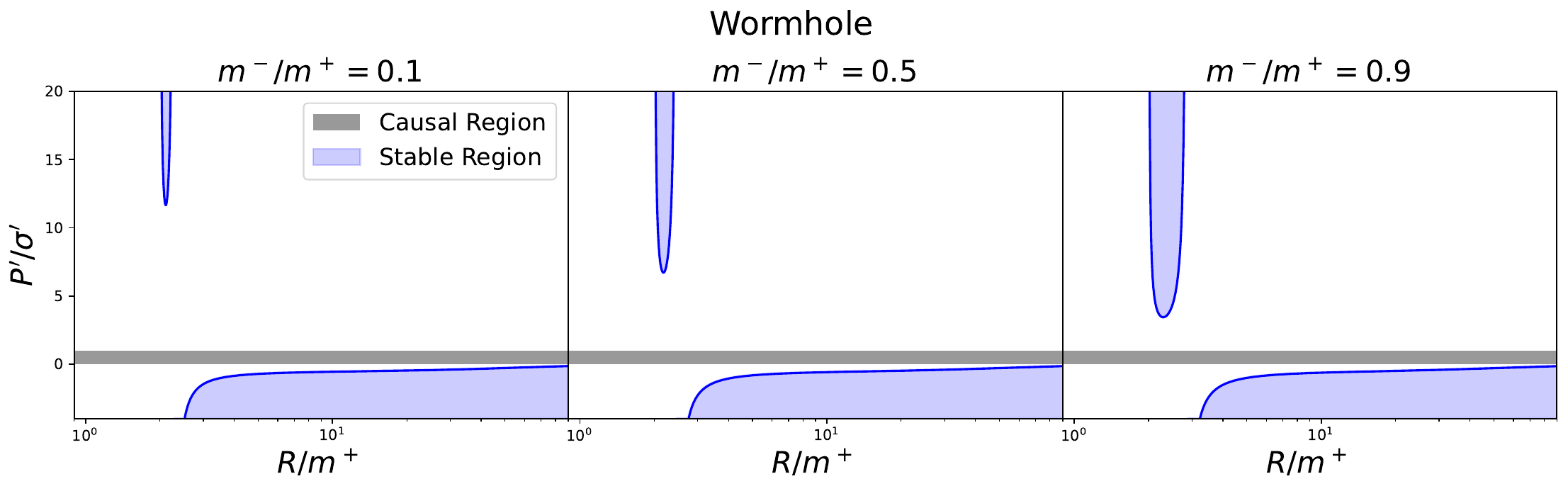}
         %\label{fig:WH}
     \end{subfigure}
     \hfill
        \caption{\textbf{Schwarzschild - anti-de Sitter -- Schwarzschild - anti-de Sitter Junction:} For the black hole, $M\neq0$ for any $\beta$ and there is no stability flip. For the wormhole, the stability region approaches the lower bound of the Causal Region as $\alpha \rightarrow \infty$ but never reaches it.}
        \label{fig:SadsSads}
\end{figure*}

%%%%%%%%FLRW%%%%%%%%
%\clearpage
\section{FLRW spacetime}
\label{sec: FLRW}

%The previous thin shell junctions, while important, have been covered in varying detail in other works (see e.g. 
%\cite{poisson1995thin}, 
%\cite{principles},
%\cite{ishak2002stability},
%\cite{Wang:2017gdp},
%\cite{Lemos:2004vs},
%\cite{Lobo:2003xd}, %\cite{Lemos:2003jb}, and
%\cite{salti2023thin}).

In this section, we turn our attention to a junction between a spacetime defined by the Friedmann-Lemaître-Robertson-Walker (FLRW) metric and a Schwarzschild or Schwarzschild - (anti-) de Sitter spacetime.
The FLRW line element is defined as
\begin{equation}
    ds^2=-c^2dt^2+a^2(t)\bigg(\frac{dr^2}{1-kr^2}+r^2d\Omega^2\bigg),
\end{equation}
where $a(t)$ is the scale factor. 

We denote the FLRW spacetime as $\mathcal{M}^-$ and the Schwarzschild spacetime as $\mathcal{M}^+$.

Unlike the previous stationary spacetimes, the FLRW spacetime is expanding. Here we follow the precedent established in the Swiss Cheese Cosmological Model, which considers spherical regions of Schwarzschild spacetime matched to an FLRW background \cite{dyer2000swisscheesecosmologicalmodel}. In these cases, it is typical to define a hypersurface that is expanding at the same rate as the FLRW so that in the FLRW frame it is stationary apart from the evolution of the hypersurface radius defined in equation (\ref{evolution eq}). The surface in the FLRW frame is defined by $r_{\rm f}=R(\tau)$. The static solution solved for here is the one for which $\dot{R}=0$.

In the Schwarzschild (or Schwarzschild - (Anti) de Sitter) frame, we define the surface as $r_{\rm s} = a(t(\tau))R(\tau)\equiv\chi(\tau)$. This allows our junction to satisfy the first Darmois condition (equation \ref{1st darmois}) as in the Swiss Cheese Model. Unlike the Swiss Cheese Model, however, we do not need to satisfy the second Darmois condition as we are using thin shells which cause a discontinuity in extrinsic curvature \cite{sahu2024singularhypersurfacesshellscosmology}.

The line element of our new hypersurface becomes
\begin{equation}
ds^2_\Sigma = -d\tau^2+a^2(t)R^2(\tau)d\Omega^2 = -d\tau^2+\chi^2(\tau)d\Omega^2.
\end{equation}

The introduction of an expanding hypersurface invalidates many of the assumptions made in previous sections. Firstly, flux term $\Xi$ is now non-zero and can be expressed as
\begin{equation}
\Xi =\pm\Bigg[\dot{r}(\rho+p){\bigg(\bigg | g^{\alpha\beta}\frac{\partial F}{\partial x^\alpha} \frac{\partial F}{\partial x^\beta}\bigg | \bigg)^{-1/2}}\Bigg],
\end{equation}
where $\rho$ and $p$ are the energy density and pressure of a perfect 4-fluid. When considering an FLRW -- Schwarzschild junction, the flux term becomes
\begin{equation}
    \Xi=\mp\frac{\rho_ma\dot{R}}{\sqrt{1-kR^2-\big(\frac{dR}{dt}\big)^2}},
\end{equation}
where $\rho_m$ is the matter energy density.

Taking $\dot{\chi} = \dot{a}R+\dot{R}a$, the conservation identity now yields
\begin{equation}
\dot{\sigma} = -2\bigg(\frac{\dot{R}}R+\frac{\dot{a}}{a}\bigg)(\sigma+P)+\frac{\rho_m\dot{R}}{\sqrt{1-kR^2}}.
\end{equation}

For a static solution $\dot{R}=0$, the identity becomes
\begin{equation}
\dot{\sigma} = -2\frac{\dot{a}}{a}(\sigma+P),
\label{flrw continutity}
\end{equation}
and the flux term does not affect the equation.

Here we break away from using prime notation to express derivatives with respect to $R$ and instead let $A'\equiv \frac{\partial A}{\partial \chi}$. Using this notation, equation (\ref{flrw continutity}) becomes
\begin{equation}
\sigma'=-\frac{2}{\chi}(\sigma+P).
\end{equation}
Note that $\sigma$ is a function of $\chi(\tau)$.

Here we let $M=-[K_{\theta\theta}]=4\pi\chi^2\sigma$. Using this, the continuity equation can finally be expressed as
\begin{equation}
\frac{P'}{\sigma'}=\frac{\dot{P}}{\dot{\sigma}}=\frac{\chi^3}{\upsilon}\bigg(\frac{M}{2\chi}\bigg)''-\frac12,
\label{flrw parameter surface}
\end{equation}
where $\upsilon\equiv3M-(M\chi)'$.

Now, we can derive the mass of the thin shell, $M$, using (equation \ref{M definition}).

For ease of computation, we define $f(r) = 1-\frac{2\mu(r)}{r}$ for the Schwarzschild and Schwarzschild - (anti-) de Sitter spacetime.

Calculating the Riemann tensor for FLRW and using equation (\ref{effective mass}), we get
\begin{equation}
    \mu(R) = \frac{\chi^3}{2}\bigg(H^2+\frac{k}{a^2}\bigg),
\label{flrw mu}
\end{equation}
where $H$ is the Hubble parameter defined as $H=\frac1a \frac{da}{dt}$ in the FLRW frame.

\begin{figure*}[t]
     \centering
     \begin{subfigure}[b]{0.95\textwidth}
         \centering
         \includegraphics[width=\textwidth]{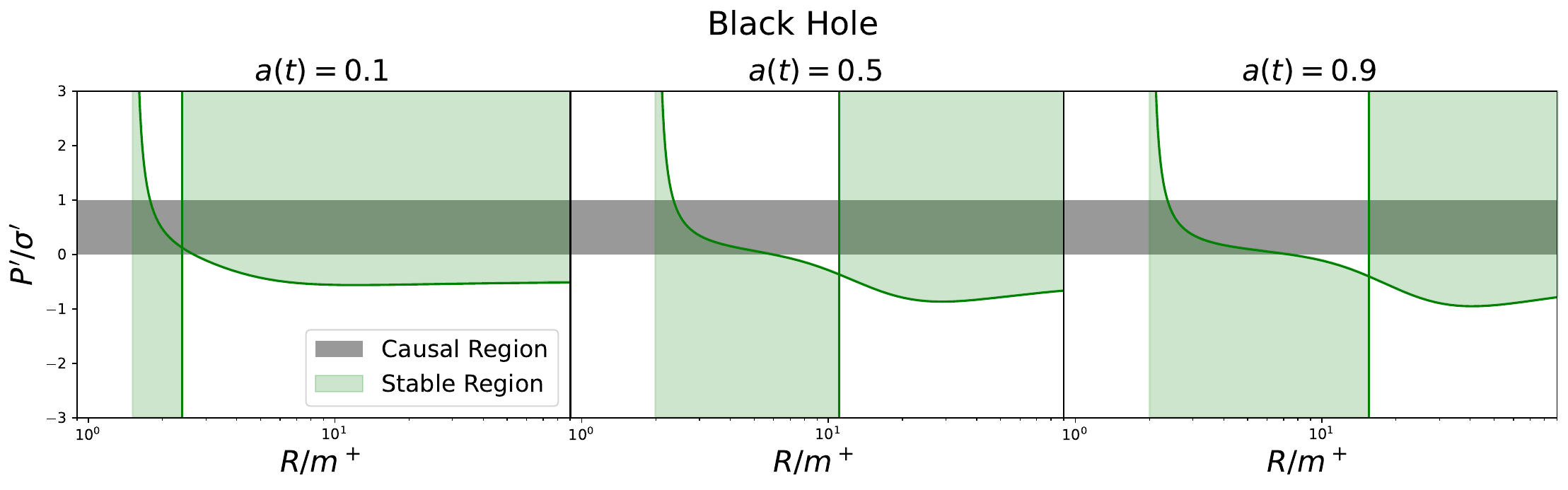}
         %\label{fig:BH}
     \end{subfigure}
     \begin{subfigure}[b]{0.95\textwidth}
         \centering
         \includegraphics[width=\textwidth]{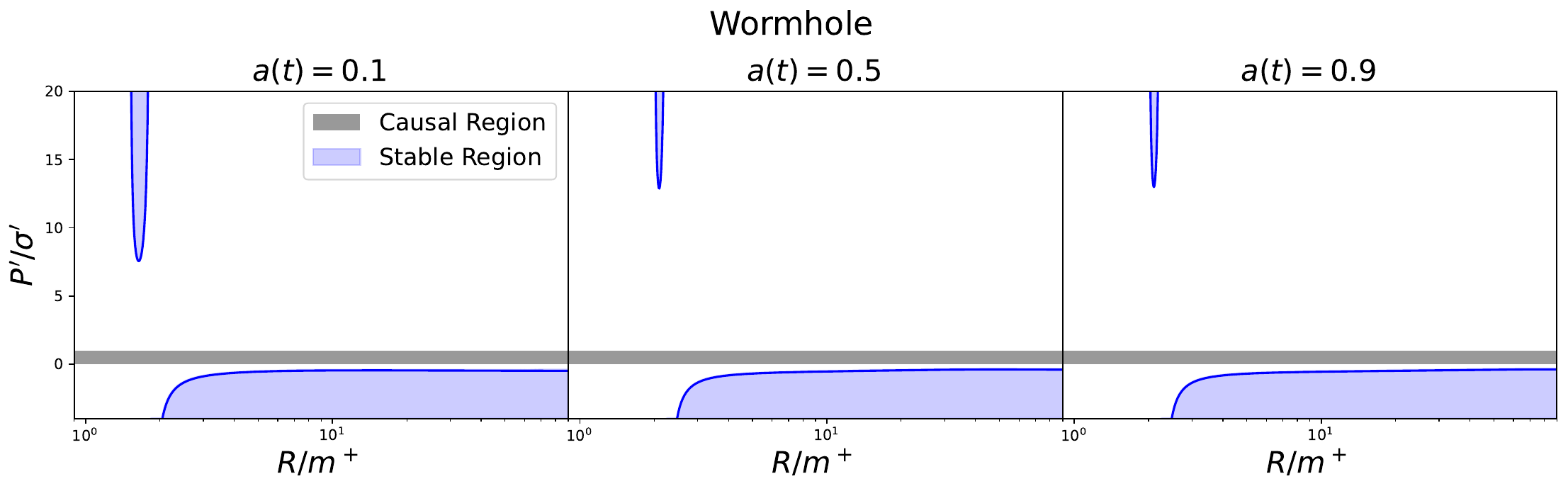}
         %\label{fig:WH}
     \end{subfigure}
     \hfill
        \caption{\textbf{FLRW (k=0) -- Schwarzschild Junction:} The black hole and wormhole graphs are similar to previous examples though does not posses a de Sitter Horizon, despite there existing a $\Lambda > 0$. It is worth noting that the radius of the event horizon decreases below $\alpha = 2$ at low $a$ due to a high value of $H^2$. For 3-D plot, see figure (\ref{fig:SF0_3d}).}
        \label{fig:FS}
\end{figure*}
\begin{figure*}[t]
     \centering
     \begin{subfigure}[b]{0.95\textwidth}
         \centering
         \includegraphics[width=\textwidth]{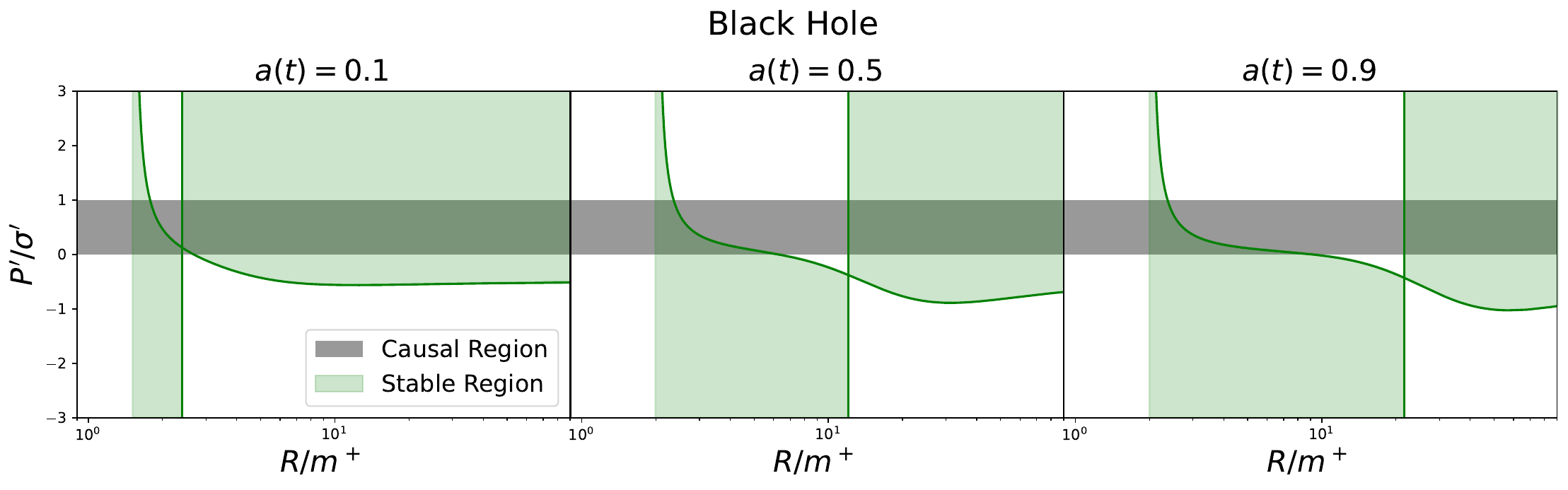}
         %\label{fig:BH}
     \end{subfigure}
     \begin{subfigure}[b]{0.95\textwidth}
         \centering
         \includegraphics[width=\textwidth]{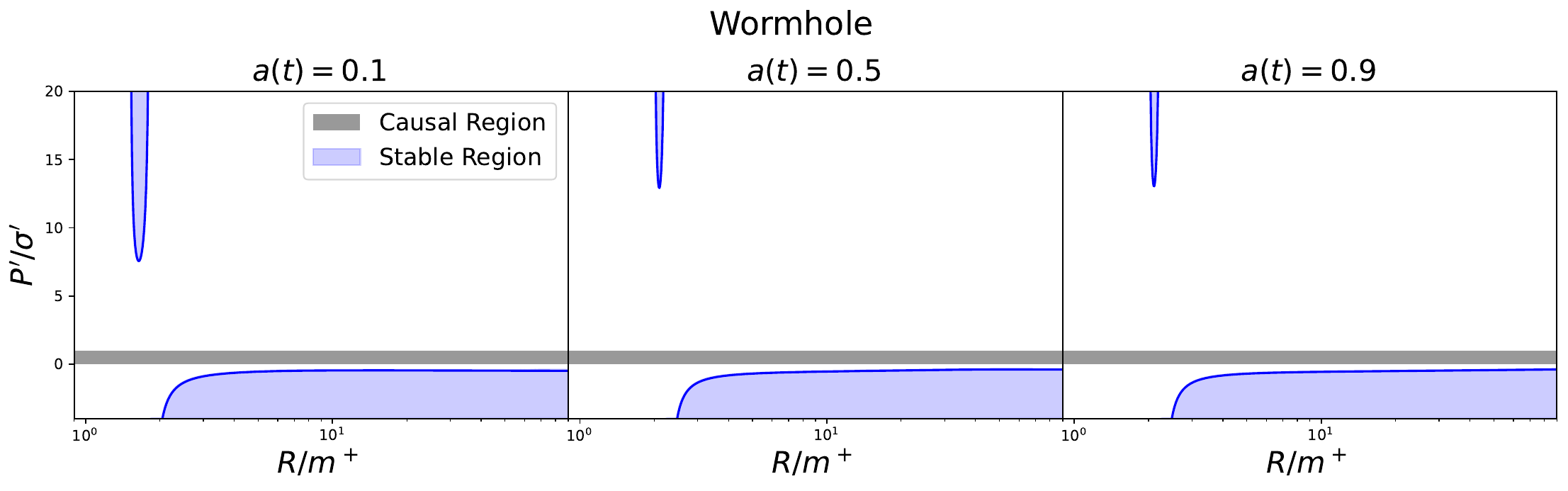}
         %\label{fig:WH}
     \end{subfigure}
     \hfill
        \caption{\textbf{FLRW (k=0) -- Schwarzschild - de Sitter Junction:}  Very similar to figure (\ref{fig:FS}). The $M=0$ line occurs at higher $\alpha$.}
        \label{fig:FSds}
\end{figure*}

The normal vector for Schwarzschild becomes
\begin{equation}
n_\gamma = \pm\bigg(f-\frac 1f\bigg(\frac{d\chi}{dt}\bigg)^2\bigg)^{-1/2} \bigg(-\frac{d\chi}{dt},1,0,0\bigg),
\end{equation}
and for the FLRW
\begin{equation}
n_\gamma = \pm\bigg(\frac{1-kR^2}{a^2}-\bigg(\frac{dR}{dt}\bigg)^2\bigg)^{-1/2} \bigg(-\frac{dR}{dt},1,0,0\bigg).
\end{equation}
The $\theta\theta$ element of the second fundamental form for both simplifies to
\begin{equation}
K_{\theta\theta} = -n_\gamma\Gamma^\gamma_{\theta \theta}.
\end{equation}

For Schwarzschild,
\begin{equation}
K_{\theta\theta}^+ = \pm \chi f^{3/2}\bigg(f^2-\bigg(\frac{d\chi}{dt}\bigg)^2\bigg)^{-1/2}.
\end{equation}

For FLRW, $K_{\theta\theta}^-$ yields a complicated form. We simplify this by taking $K_{\theta \theta}$ at the static solution $\dot{R} = \frac{dR}{dt}\frac{dt}{d\tau}=0 \Rightarrow \frac{dR}{dt} = 0$ if $\frac{dt}{d\tau} \neq 0$.

Thus,
\begin{equation}
K_{\theta\theta}^- = \pm\chi\sqrt{1-kR^2}.
\end{equation}

From the first Darmois condition (equation \ref{1st darmois}) and using the definition of the induced metric we can find an equation for $\frac{dt}{d\tau}$ for both spacetimes.

For Schwarzschild:
\begin{equation}
\bigg(\frac{dt}{d\tau}\bigg)^2=\frac1{f^2}(f+\dot{\chi}^2),
\end{equation}
and for FLRW:
\begin{equation}
\bigg(\frac{dt}{d\tau}\bigg)^2=\frac{a^2}{1-kR^2}\dot{R}^2+1.
\end{equation}

Plugging into 
the Schwarzschild $K_{\theta\theta}$ yields
\begin{eqnarray}\nonumber
K_{\theta\theta}^+ &=& \chi\sqrt{f+\dot{\chi}}, \\ &=& \chi\sqrt{1-\frac{2\mu(\chi)}{\chi}+\dot{\chi}^2}.
\end{eqnarray}

For a static solution, in the FLRW frame
\begin{equation}
aH = \frac{da}{dt} = \dot{a}\frac{dt}{d\tau} = \dot{a}.
\end{equation}
This implies
\begin{equation}
    \dot{\chi}=\dot{a}R=H\chi.
\end{equation}
Utilizing the above and (equation \ref{flrw mu}) we can express the $K_{\theta\theta}$ for the FLRW spacetime in the same manner,
\begin{equation}
K_{\theta\theta}^- = \chi\sqrt{1-\frac{2\mu(\chi)}{\chi}+\dot{\chi}^2}.
\end{equation}

Using (equation \ref{M definition}) we can derive a new equation of motion,
\begin{equation}
\dot{\chi}^2=\bigg(\frac{[\mu]}{M}\bigg)^2+\frac{2\bar{\mu}}{\chi}+\bigg(\frac{M}{2\chi}\bigg)^2-1=-V(\chi(R)),
\end{equation}
which is in the same form as before.

Though, at the static solution $V =H\chi\neq 0$ generally. $V'$ may be be chosen to be $0$ for an equilibrium point. Thus, $V''>0$ again yields stable equilibria. Combining this with the transparency condition (equation \ref{flrw parameter surface}) and the fact that $V'=0$ at this point, gives stability conditions of

\begin{eqnarray}
    \frac{P'}{\sigma'} &<& \frac{\chi^3}{\upsilon}\bigg(\frac{M}{2\chi}\bigg)''-\frac12; \quad M\upsilon>0 \rm,
    \label{F stability condition 1}
    \\ 
    \frac{P'}{\sigma'} &>& \frac{\chi^3}{\upsilon}\bigg(\frac{M}{2\chi}\bigg)''-\frac12; \quad M\upsilon<0.
    \label{F stability condition 2}
\end{eqnarray}

Finally, we can express $M$ in a static case where $\dot{\chi}=H\chi$ as
\begin{equation}
    M = w\chi\sqrt{1-k\frac{\chi^2}{a^2}}-\chi\sqrt{1-\frac{2m}{\chi}-\bigg(\frac{\Lambda^+}{3}-H^2\bigg)\chi^2}.
\end{equation}

It is now clear that each new $K_{\theta\theta}$ takes on the same form as before but now has an additional positive term $\dot{\chi}=H\chi$ that incorporates the expansion.
This positive term acts like a negative mass. For example, in the Schwarzschild spacetime, the expansion term allows an event horizon radius smaller than the traditional Schwarzschild radius.

$M$ may also be recast into a form analogous to equation (\ref{Sch M}) as
\begin{equation}
    M = w\chi\sqrt{1-\frac{2m^-}{\chi}-\frac{\varepsilon^- \chi^2}{3}} - \chi\sqrt{1-\frac{2m^+}{\chi}-\frac{\varepsilon^+ \chi^2}{3}}.
\end{equation}
Instead of using $\Lambda$, we define $\varepsilon/3$ to be the coefficient of the $\chi^2$ term in each $K_{\theta\theta}$.

\begin{figure*}[t]
     \centering
     \begin{subfigure}[b]{0.95\textwidth}
         \centering
         \includegraphics[width=\textwidth]{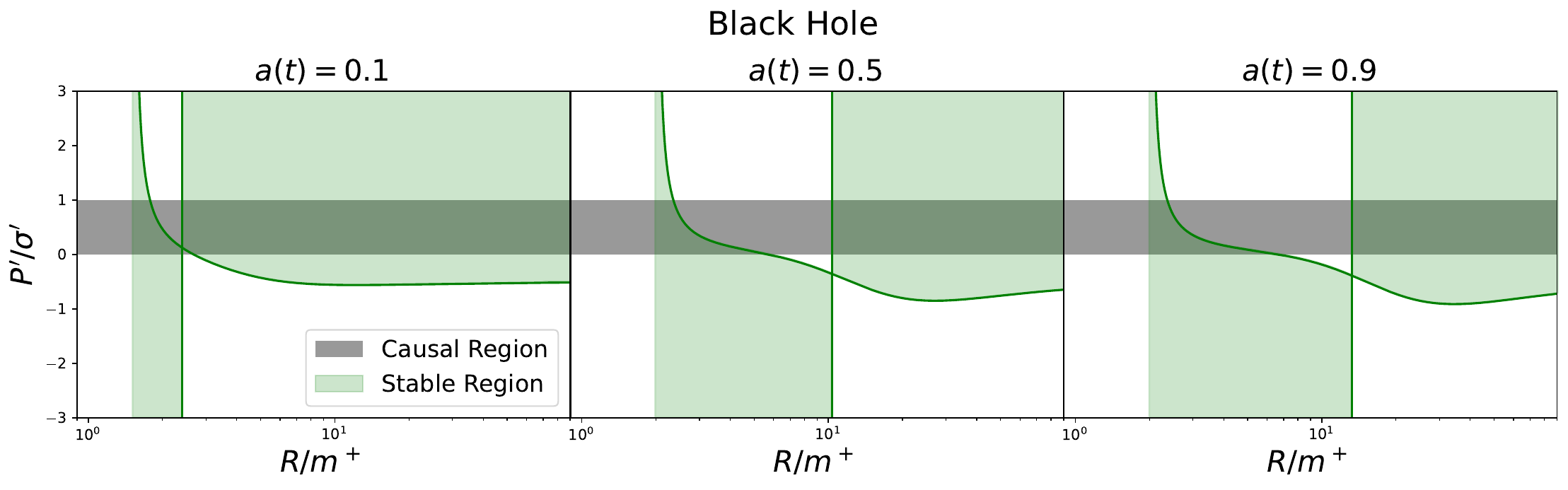}
         %\label{fig:BH}
     \end{subfigure}
     \begin{subfigure}[b]{0.95\textwidth}
         \centering
         \includegraphics[width=\textwidth]{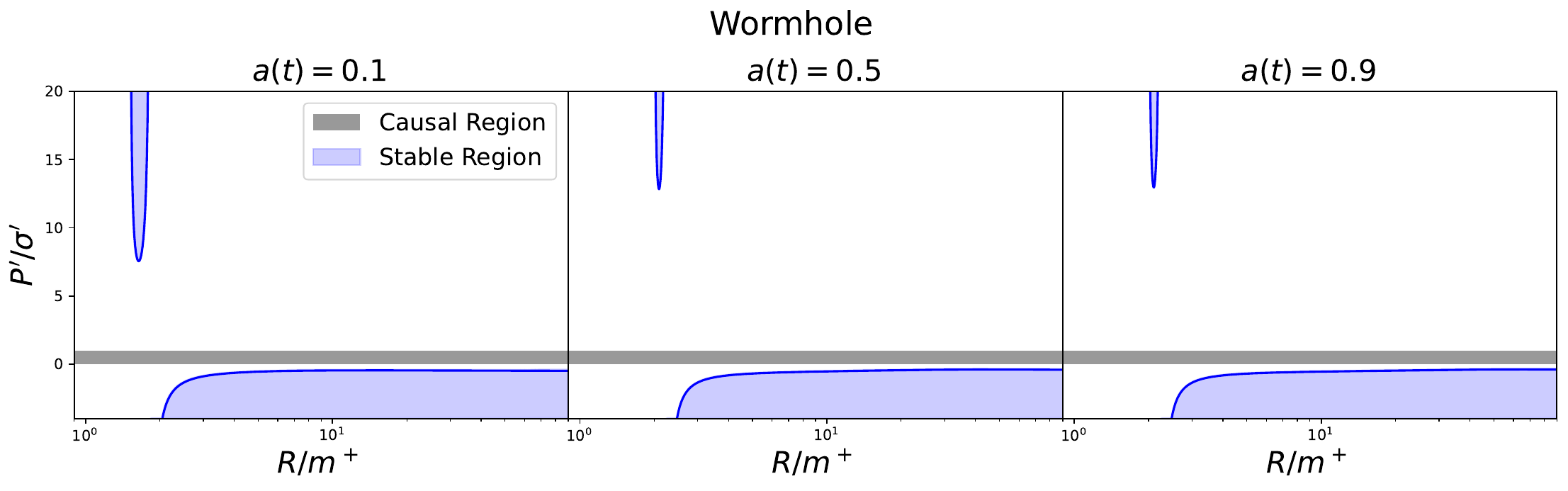}
         %\label{fig:WH}
     \end{subfigure}
     \hfill
        \caption{\textbf{FLRW (k=0) -- Schwarzschild - anti-de Sitter Junction:} Very similar to figure (\ref{fig:FS}). The $M=0$ line occurs at lower $\alpha$.}
        \label{fig:FSads}
\end{figure*}

We may now use the same asymptote conditions as in Section \ref{sec:Schwarzschild} just substituting $\varepsilon$ for $\Lambda$.

The first Friedmann Equation is
\begin{equation}
    H^2=\frac{8\pi}{3}\rho-\frac{k}{a^2}+\frac{\Lambda^-}{3}.
\end{equation}

For Schwarzschild or Schwarzschild - (anti-) de Sitter, we can define $\varepsilon$ as
\begin{equation}
    \varepsilon^+ = \Lambda^+ - 3H^2 = \frac{3k}{a^2} -8\pi \rho + [\Lambda],
\end{equation}
where $\Lambda^+$ is the cosmological constant of the Schwarzschild - (anti-) de Sitter and $\Lambda^-$ is the cosmological constant of the FLRW. For Schwarzschild, $\Lambda^+ = 0$. 

For FLRW $\varepsilon$ is simply
\begin{equation}
\varepsilon^- = \frac{3k}{a^2}.
\end{equation}

For all Schwarzschild spacetimes, $m^+ = m$, where $m$ is the Schwarzschild mass and is constant. For FLRW, $m^- = 0$. 

The condition for the existence of a de Sitter horizon becomes $\varepsilon^+>0$ or $\varepsilon^->0$ which can be expressed as
\begin{equation}
[\Lambda] > 8\pi\rho - \frac{3k}{a^2},
\label{FLRW desitter horizon 1}
\end{equation}
or
\begin{equation}
3k > 0 \Rightarrow k = +1,
\label{FLRW desitter horizon 2}
\end{equation}
since $k$ must be either -1, 0, or +1.

Lastly, the asymptotic conditions (equations \ref{Lam < 0 asymptote condition}, \ref{asymptote condition}) are re-expressed for $\upsilon=0$ as
\begin{equation}
\frac{[\varepsilon]}{[m]}>0,
\label{lam<0 asymtote condition ep}
\end{equation}
\begin{equation}
\frac{[\varepsilon]}{[m]}>\frac{-2\varepsilon_{\rm max}^\frac32}{\sqrt3}.
\label{asymtote condition ep}
\end{equation}
% It is important to note that since $\dot{{\chi}}=H\chi\neq0$, while the presence of an asymptote at $\upsilon=0$ is still predicted by the conditions, $M=0$ now also results in an asymptote when $H^2\neq0$. When $H^2=0$ this asymptote disappears and $M=0$ results in a stability flip as in the previous section.

\begin{figure*}[t]
     \centering
     \begin{subfigure}[b]{0.95\textwidth}
         \centering
         \includegraphics[width=\textwidth]{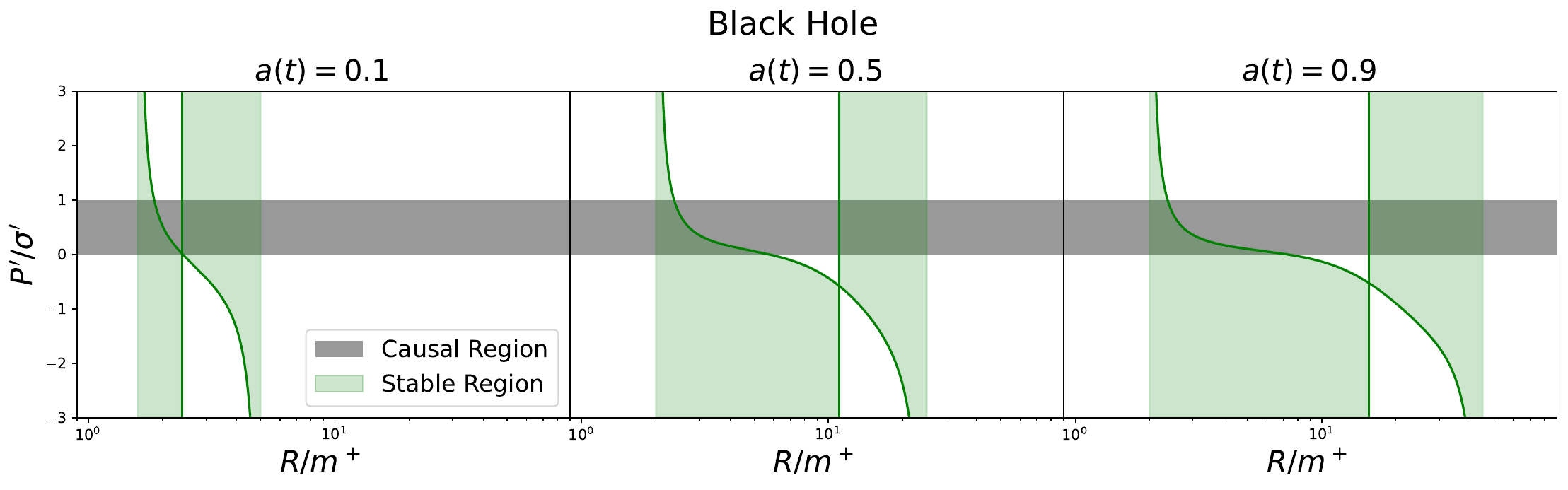}
         %\label{fig:BH}
     \end{subfigure}
     \begin{subfigure}[b]{0.95\textwidth}
         \centering
         \includegraphics[width=\textwidth]{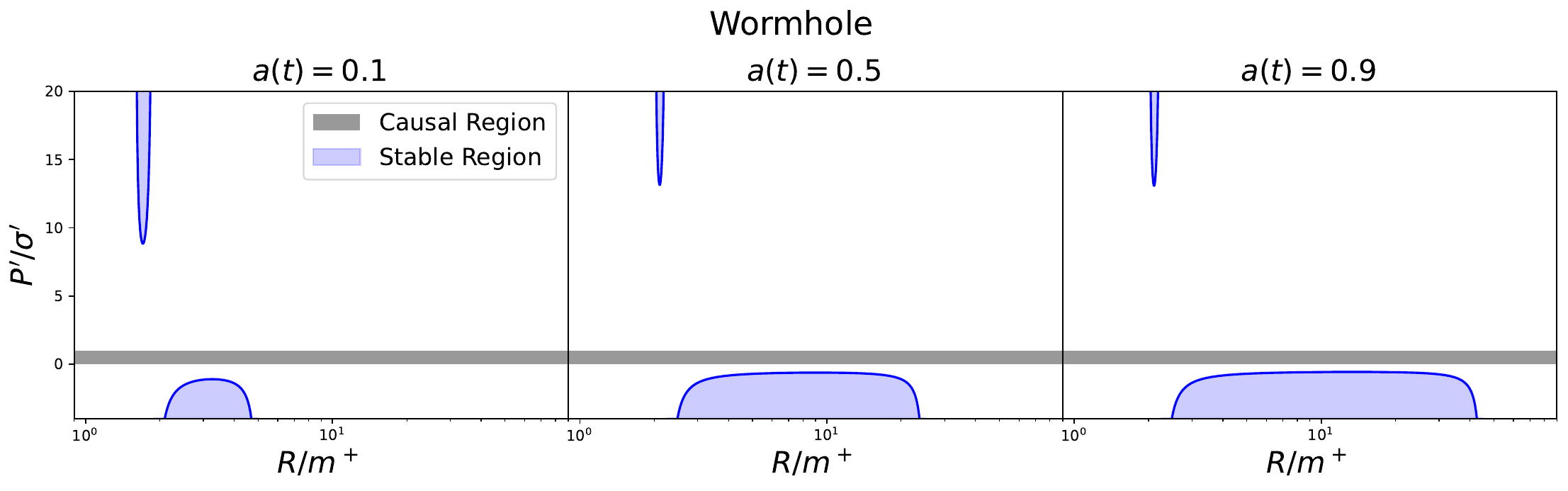}
         %\label{fig:WH}
     \end{subfigure}
     \hfill
        \caption{\textbf{FLRW (k=+1) -- Schwarzschild Junction:} For $k\neq0$ we choose $m=0.02$. It is important to note that $k>0$ gives the non physical result of $H^2<0$ for $0.420 \lesssim a(t)\lesssim 0.823$. A horizon caused by the positive curvature exists at high $\chi$ for both black hole and wormhole. At larger $m$ values the asymptote condition can be fulfilled but commonly this occurs only when $H^2<0$. For 3-D plot, see figure (\ref{fig:SF+1_3d}).}
        \label{fig:F+1S}
\end{figure*}

\section{Results \label{sec:results}}
\par\noindent

In the following, we turn our attention to the plotting of regions of stability in three-dimensional parameter space. Overall qualitative features of the observed stability regions are discussed.

\subsection{Plotting \label{sec:plotting}}
\par\noindent

When utilizing the results of the previous sections, it is helpful to make a coordinate transformation for easier plotting. We consider $\alpha = R/m^+$ $\beta = m^-/m^+$, and define $C^\pm = \Lambda^\pm m^{+2}$.

For Schwarzschild or Schwarzschild - (anti-) de Sitter, we have
\begin{equation}
M = w m^+\alpha\sqrt{1-2\frac{\beta}{\alpha}-\frac{C^-\alpha^2}{3}}-m^+\alpha\sqrt{1-2\frac{1}{\alpha}-\frac{C^+\alpha^2}{3}}.
\end{equation}

We can now view the parameter space by plotting $P'/\sigma'$ as a function of $\alpha$ and $\beta$. Since $m^+$ is not a function of $R$, when plugging into equations (\ref{stability condition 1}) and (\ref{stability condition 2}), it cancels and have no effect on the final stability surface.
Following a similar approach as \cite{ishak2002stability}, we choose $|C^\pm|=|\Lambda^\pm m^{+2}| \sim 10^{-3}$. This assumption gives a high value for this product, but is useful for qualitative plotting.

Applying the substitutions to the asymptotic conditions given by equations (\ref{Lam < 0 asymptote condition}) and (\ref{asymptote condition}) yields
\begin{equation}
    [C] > 0, 
\end{equation}
since $1-\beta > 0$, and
\begin{equation}
    [C] > \frac{-2 C_{\rm max}^{\frac32}}{\sqrt3} (1-\beta).
\end{equation}
It can also be seen that $\alpha_{\rm \rm dS}\approx\sqrt{3/C_{\rm max}}$.

For the junctions between FLRW and Schwarzschild, we consider $\alpha = \chi/m$ and $C^\pm = \Lambda^\pm m^2$. Note that $\beta =m^-/m=0$ in this case as $m^-=0$.

We choose $C^+ = \pm10^{-3}$ for Schwarzschild - de Sitter or Schwarzschild - anti-de Sitter and $C^-=+10^{-3}$ always. Thus, $[C] \leq 0$.

For the FLRW -- Schwarzschild junctions, using the continuity equation of the FLRW spacetime, the evolution of $\rho$ can be expressed as $\left(\frac{\rho^0}{a^3} \right)$ where $\rho^0$ is a constant. We define $D \equiv 8\pi m^2\rho^0$ to keep dimensional consistency with $C$. Using our assumption for the magnitude of $C$ and the value of $\rho_\Lambda$ we find that for the present universe ($a(t)=1$) $D/C=\rho^0/\rho_\Lambda=\Omega_m^0/\Omega_\Lambda\approx0.3/0.7$.
Thus, $D \sim 4.23\times10^{-4}$. From this, $M$ is as follows,
\begin{eqnarray}\nonumber
M &=& wm\alpha\sqrt{1-\frac{km^2}{a^2}\alpha^2}\\
&& -m\alpha\sqrt{1-\frac{2}{\alpha} - \bigg(\frac{3km^2}{a^2} - \frac{D}{a^3} +[C]\bigg)\frac{\alpha^2}{3}}.
\end{eqnarray}

Again, since $m$ is constant with respect to $\chi$, it cancels when plugged into equations (\ref{F stability condition 1}) and (\ref{F stability condition 2}).
However, the contribution from curvature $k$ is affected by the $m^2$ factor which cannot be ignored. A value for this $m$ must be chosen when plotting $k\neq0$.

\begin{figure*}[t]
     \centering
     \begin{subfigure}[b]{0.95\textwidth}
         \centering
         \includegraphics[width=\textwidth]{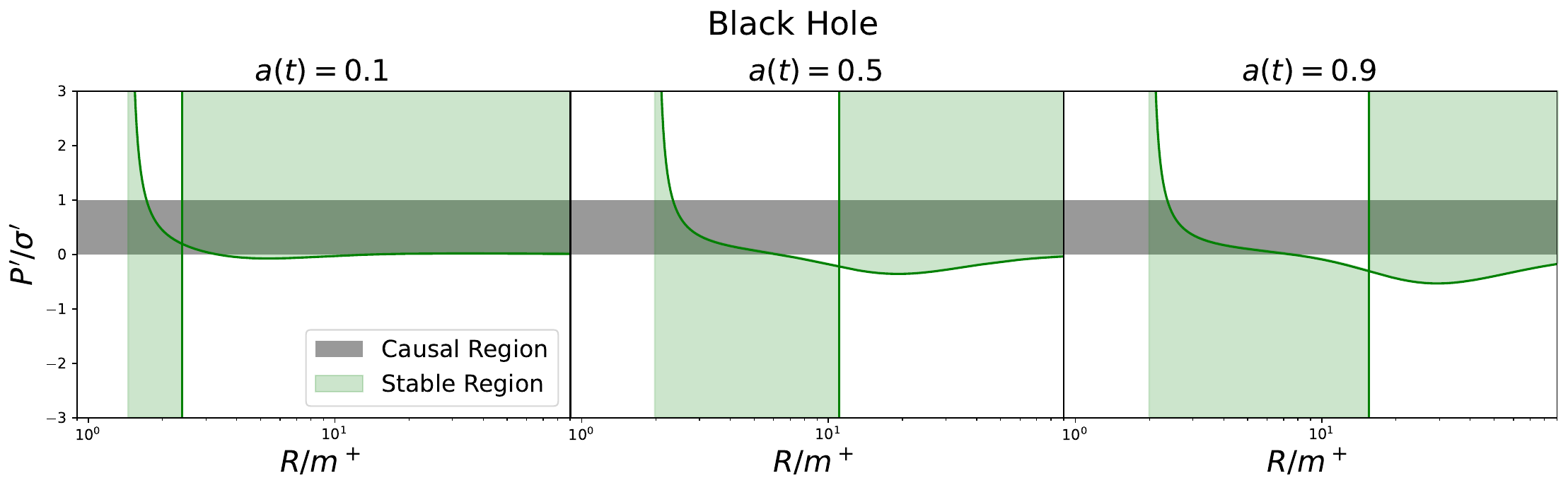}
         %\label{fig:BH}
     \end{subfigure}
     \begin{subfigure}[b]{0.95\textwidth}
         \centering
         \includegraphics[width=\textwidth]{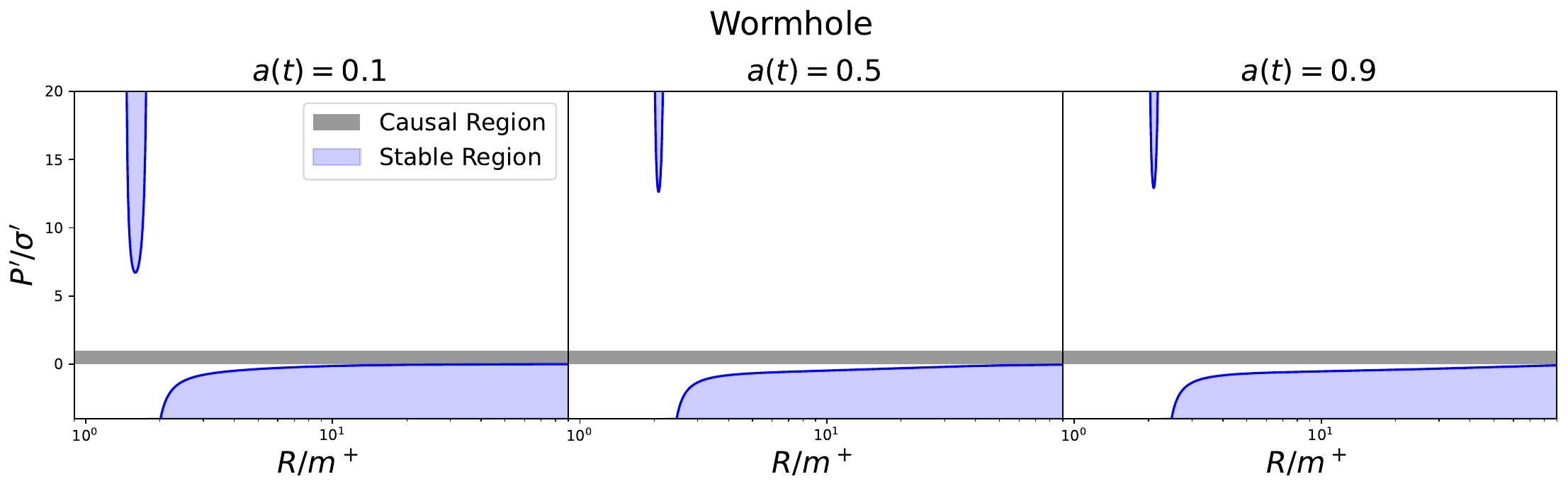}
         %\label{fig:WH}
     \end{subfigure}
     \hfill
        \caption{\textbf{FLRW (k=-1) -- Schwarzschild Junction:} 
        There is no horizon present at high $\chi$ for either the black hole or the wormhole. As in figure (\ref{fig:SadsSads}), the stability region of the wormhole approaches the lower bound of the Causal Region as $\alpha \rightarrow \infty$. For 3-D plot, see figure (\ref{fig:SF-1_3d}).}
        \label{fig:F-1S}
\end{figure*}

Applying the substitutions to the conditions for the existence of a de Sitter Horizon (equations \ref{FLRW desitter horizon 1} and \ref{FLRW desitter horizon 2}) gives
\begin{equation}
[C] > \frac{D}{a^3}-\frac{3km^2}{a^2}.
\label{1st horizon}
\end{equation}
or
\begin{equation}
k=+1.
\end{equation}
However, due the fact that $[C]\leq0$, equation (\ref{1st horizon}) is never satisfied when $k\neq+1$.

Applying the substitutions to the asymptotic conditions in equation (\ref{lam<0 asymtote condition ep}) yields
\begin{equation}
   [C] > \frac{D}{a^3},
\end{equation}
which is never satisfied for $[C]\leq0$.

When using equation (\ref{asymtote condition ep}) we assume $\varepsilon_{\rm max}=\varepsilon^-=3km^2$ as $D>0$ and $[C]<0$ cause $\varepsilon^->\varepsilon^+$. This condition is only relevant if $\varepsilon^->0$ which implies the $k=+1$ case. So,
\begin{equation}
[C]>\frac{D-6m^3}{a^3},
\end{equation}
which implies if $D-6m^3>0$ there can be no asymptote at any $a(t)$ for $[C]\leq0$.
Additionally, $\alpha_{\rm dS}\approx\frac{1}{m}$ when $k=+1$. Such a horizon is nonexistent for $k\neq+1$.

Plotted examples of parameter spaces for Schwarzschild and Schwarzschild - (anti-) de Sitter junctions are given in figures (\ref{fig:SS}-\ref{fig:SadsSads}). Figures (\ref{fig:FS}-\ref{fig:F-1S}) detail parameter spaces of FLRW junctions with Schwarzschild and Schwarzschild - (anti-) de Sitter. Three dimension plots can be found in Appendix \ref{sec:3d plots} for select junctions.

%Table 1
\begin{table*}

% \begin{ruledtabular}
\begin{tabular}{|c|c|c|c|c|}
\hline \hline
Type & Asymptote at $\Upsilon = 0$ & Limit as $R \rightarrow \infty$ & Stability in Causal Region & Figure \\\hline
Sch -- Sch & No & Convergent & Yes & \ref{fig:SS} \\
Sch -- Sch-deSit & Yes & Divergent & Yes & \ref{fig:SSds} \\
Sch -- Sch-Anti deSit & No & Convergent & Yes & \ref{fig:SSads} \\
Sch-deSit -- Sch & No & Divergent & Yes & \ref{fig:SdsS}\\
Sch-deSit -- Sch-deSit & Yes & Divergent & Yes & \ref{fig:SdsSds}\\
Sch-deSit -- Sch-Anti deSit & No & Divergent & Yes & -\\
Sch-Anti deSit -- Sch & Yes & Convergent & Yes & \ref{fig:SadsS}\\
Sch -Anti deSit -- Sch-deSit & Yes & Divergent & Yes & -\\
Sch-Anti deSit -- Sch-Anti deSit & No & Convergent & Yes & \ref{fig:SadsSads}\\
FLRW (k=0) -- Sch & No & Convergent & Yes & \ref{fig:FS}\\
FLRW (k=0) -- Sch-deSit & No & Convergent & Yes & \ref{fig:FSds}\\
FLRW (k=0) -- Sch-Anti deSit & No & Convergent & Yes & \ref{fig:FSads}\\
FLRW (k=+1) -- Sch & No & Divergent & Yes & \ref{fig:F+1S}\\
FLRW (k=-1) -- Sch & No & Convergent & Yes & \ref{fig:F-1S}
\\
\hline \hline
\end{tabular}
\caption{\label{tab:table1} Summary of the notable features of the parameter spaces that explain the stability of a black hole in the Causal Region.}
% Notable Features of Black hole Parameter Spaces.
% \end{ruledtabular}
\end{table*}

%Table 2
\begin{table*}
% \begin{ruledtabular}
\begin{tabular}{|c|c|c|c|c|}
\hline \hline
Type & Asymptote at $\Upsilon = 0$ & Limit as $R \rightarrow \infty$ & Stability in Causal Region & Figure \\\hline
Sch -- Sch & Yes & Convergent & No & \ref{fig:SS}\\
Sch -- Sch-deSit & Yes & Divergent & No & \ref{fig:SSds}\\
Sch -- Sch-Anti deSit & Yes & Convergent & No & \ref{fig:SSads} \\
Sch-deSit -- Sch & Yes & Divergent & No & \ref{fig:SdsS} \\
Sch-deSit -- Sch-deSit & Yes & Divergent & No & \ref{fig:SdsSds} \\
Sch-deSit -- Sch-Anti deSit & Yes & Divergent & No & - \\
Sch-Anti deSit -- Sch & Yes & Convergent & No & \ref{fig:SadsS}\\
Sch -Anti deSit -- Sch-deSit & Yes & Divergent & No & -\\
Sch-Anti deSit -- Sch-Anti deSit & Yes & Convergent & No & \ref{fig:SadsSads} \\
FLRW (k=0) -- Sch & Yes & Convergent & No & \ref{fig:FS}\\
FLRW (k=0) -- Sch-deSit & Yes & Convergent & No & \ref{fig:FSds} \\
FLRW (k=0) -- Sch-Anti deSit & Yes & Convergent & No & \ref{fig:FSads} \\
FLRW (k=+1) -- FLRW & Yes & Divergent & No & \ref{fig:F+1S} \\
FLRW (k=-1) -- FLRW & Yes & Convergent & No & \ref{fig:F-1S}
\\
\hline \hline
\end{tabular}
\caption{\label{tab:table2} Summary of the notable features of the parameter spaces that explain the stability of a wormhole in the Causal Region.
% Notable Features of Wormhole parameter Spaces.
}
% \end{ruledtabular}
\end{table*}

\subsection{Discussion \label{sec:discussion}}
\par\noindent

From the results seen in the figures throughout and summarized in Tables \ref{tab:table1} and \ref{tab:table2}, we can see that the predictions made in Section \ref{sec:Schwarzschild} and \ref{sec: FLRW} are met. All (FLRW included) wormhole junctions possess an asymptote at
$\Upsilon = 0$. The location of the asymptote can vary, but in all cases it divides the parameter space into two disjoint stability regions. The stability region at lower $R$ or $\chi$ is bound by a minimum $\frac{P'}{\sigma'}>0$. The stability region at higher $R$ and is bound by a maximum $\frac{P'}{\sigma'}<0$. In all cases observed these stability regions do not intersect with the Causal Region, $0\leq\frac{P'}{\sigma'}<1$.

All (FLRW included) black hole junctions either possess an asymptote
or an $M = 0$ stability flip with the exception of the simple
Schwarzschild -- Schwarzschild and Schwarzschild - anti-de Sitter -- Schwarzschild - anti-de Sitter. Of
those that have some form of stability flip, (either due
to $M=0$ or $\Upsilon=0$) the stability is defined by two regions. The region at low $R$ or $\chi$ and is defined by a maximum $\frac{P'}{\sigma'}$. Region B is at higher R and is defined by a minimum $\frac{P'}{\sigma'}$. The maximum is above $\frac{P'}{\sigma'}=0$, causing an intersection of the stability regions with the Causal region. However, such an intersection does not exist for every combination of $\alpha$ and $\beta$ or $\alpha$ and $a(t)$.

In our plotting, since $m^+ \geq m^-$, the swapping of spacetimes $\mathcal{M}^+$ and $\mathcal{M}^-$ has a noticeable effect on the stability regions. However, if one also changes the sign of $\Lambda$ in this swap, the qualitative features of the plot are preserved. For example, Schwarzschild Anti-de Sitter -- Schwarzschild (figure \ref{fig:SadsS}) and Schwarzschild -- Schwarzschild - de Sitter (figure \ref{fig:SSds}) produce similar plots.

FLRW junctions act similar to junctions with only Schwarzschild, Schwarzschild - (anti-) de Sitter but lack asymptotes for black holes. Additionally, due to the positive expansion term $H^2\chi^2$, the event horizon approaches zero for low $a(t)$. A cosmological horizon is only permitted for $[C]>0$ (which is beyond the scope of this paper) or $k=+1$.

Each black hole and wormhole with and without an FLRW spacetime can be successfully grouped into the 4 categories, as can be seen in Tables \ref{tab:table1} and \ref{tab:table2}.

\section{Conclusion \label{sec:Conclusion}}
\par\noindent

We performed a stability analysis of thin shell wormholes and black holes constructed from combinations of Schwarzschild, Schwarzschild - de Sitter,  Schwarzschild - anti-de Sitter, and Friedmann-Lemaître-Robertson-Walker (FLRW) spacetimes. We provide a taxonomy of these combinations while emphasizing some common landscape characteristics with the localization of stability and causality regions. 

% As previously stated, 
We found no parameter combinations which yield a stable wormhole when restricting $0\leq\frac{P'}{\sigma'}<1$. The boundaries of the stability regions consistently lie outside the causal region, either at very high values or at negative ones. This conclusion 
% has been 
was echoed in previous works \cite{poisson1995thin, ishak2002stability, Lobo:2005zu, Lobo_2005}. Our investigation of junctions containing the FLRW spacetime have yielded the same results.

One may interpret $P'/\sigma'$ as the square of the sound speed of perturbation, in which case it would be necessary to restrict its values to the causal region where the sound speed could not exceed the speed of light and be restricted to the reals. In this case, one would tend to deny the stability of the considered wormholes in the physical universe. However, as it was cautioned in \cite{poisson1995thin, Lobo:2005zu, Lobo_2005}, it may not be fully justified to entirely rule stability out. Indeed, the geometry of a wormhole already requires a violation of the null energy condition \cite{poisson1995thin}. Such unusual conditions could allow for $P'/\sigma'$ to take on values outside of the causal range. It has been noted that negative $P'/\sigma'$ values can occur with the Casimir Effect and the False Vacuum \cite{poisson1995thin}. Unfortunately, until a detailed model of exotic matter is formulated, it remains unclear if such values should be excluded from the discussion or not. It is also of note that possible stability in the causal region has been noted in rotating (BTZ metric) wormholes for sufficient angular momentum \cite{Tsukamoto:2018lsg}. Thus, these wormholes do not require a relaxation of the $0<P'/\sigma'<1$ condition to be stable.

%using thin shell formalism, we have evaluated the shape, location, and presence of stability regions of junctions (both black holes and wormholes) of different combinations of Schwarzschild, Schwarzschild - de Sitter, and Schwarzschild - anti-de Sitter spacetimes of different mass ratios and radii. We have also formulated an approach for considering junctions with an expanding spacetime governed by the FLRW metric and evaluated the stability regions produced by such a formulation. 

Finally, we have categorized the different junctions into taxonomic groups and depicted the mathematical conditions under which these groups are formed. While we found stability in the causal region only for black holes configurations, it remains possible that stable wormholes could exist, though only in extraordinary conditions which may not be allowed from the point of the view of semi-classical GR. This is still found to be the case even when FLRW spacetimes are used in the configurations.

%Table 1

%Table 3
% \begin{table*}
% \caption{\label{tab:table3}Classification of Black hole Parameter Spaces.}
% \begin{ruledtabular}
% \begin{tabular}{|c|c|c|c|}
% Asymptotic Convergent & Asymptotic Divergent & Continuous Convergent & Continuous Divergent \\\hline
% Sch-Anti deSit|Sch & Sch|Sch-deSit & Sch|Sch & Sch-deSit|Sch \\
%  & Sch-deSit|Sch-deSit & Sch|Sch-Anti deSit & Sch-deSit|Sch-Anti deSit \\
%  & Sch-Anti deSit|Sch-deSit & Sch-Anti deSit|Sch-Anti deSit & FLRW k=+1|Sch\\
%  &  FLRW k=+1|Sch \footnote{For high $a(t)$ and high $m_s$} & FLRW k=0|Sch & \\
%   & & FLRW k=0|Sch-deSit & \\
%  & & FLRW k=0|Sch-Anti deSit &  \\
%   & & FLRW k=-1 &  \\

% \end{tabular}
% \end{ruledtabular}
% \end{table*}

% %Table 4
% \begin{table*}[hbt!]
% \caption{\label{tab:table4}Classification of Wormhole Parameter Spaces.}
% \begin{ruledtabular}
% \begin{tabular}{|c|c|c|c|}
% Asymptotic Convergent & Asymptotic Divergent & Continuous Convergent & Continuous Divergent\\\hline
% Sch|Sch & Sch|Sch-deSit & & \\
% Sch|Sch-Anti deSit & Sch-deSit|Sch & & \\
% Sch-Anti deSit|Sch & Sch-deSit|Sch-deSit & & \\
% Sch-Anti deSit|Sch-Anti deSit & Sch-deSit|Sch-Anti deSit & & \\
% FLRW k=0|Sch & Sch-Anti deSit|Sch-deSit & & \\
% FLRW k=0|Sch-deSit & FLRW k=1|Sch-deSit & & \\
% FLRW k=0|Sch-Anti deSit & & & \\

% \end{tabular}
% \end{ruledtabular}
% \end{table*}

\section*{Acknowledgements}
\par\noindent

We thank Francisco Lobo for useful comments on the manuscript. MI acknowledges that this material is based upon work supported in part by the Department of Energy, Office of Science, under Award Number DE-SC0022184 and also in part by the U.S. National Science Foundation under grant AST2327245.

\bibliography{Bibliography}

%apssamp
%\printbibliography
% {
%     %\small

% % \bibliographystyle{ieeenat_fullname}
% % \bibliographystyle{plain}
% \bibliographystyle{apsrev4-2} 
% \bibliography{Sources}
% % \nocite{*}
% }

% \newpage
% \onecolumn
% \begin{onecolumn}
  % Your single-column content here

\newpage

\onecolumngrid
\appendix
\section{3-D Plots}
\label{sec:3d plots}
\par
\begin{figure*}[hbt!]
     \centering
     \begin{subfigure}[b]{0.47\textwidth}
         \centering
        \includegraphics[width=\textwidth]{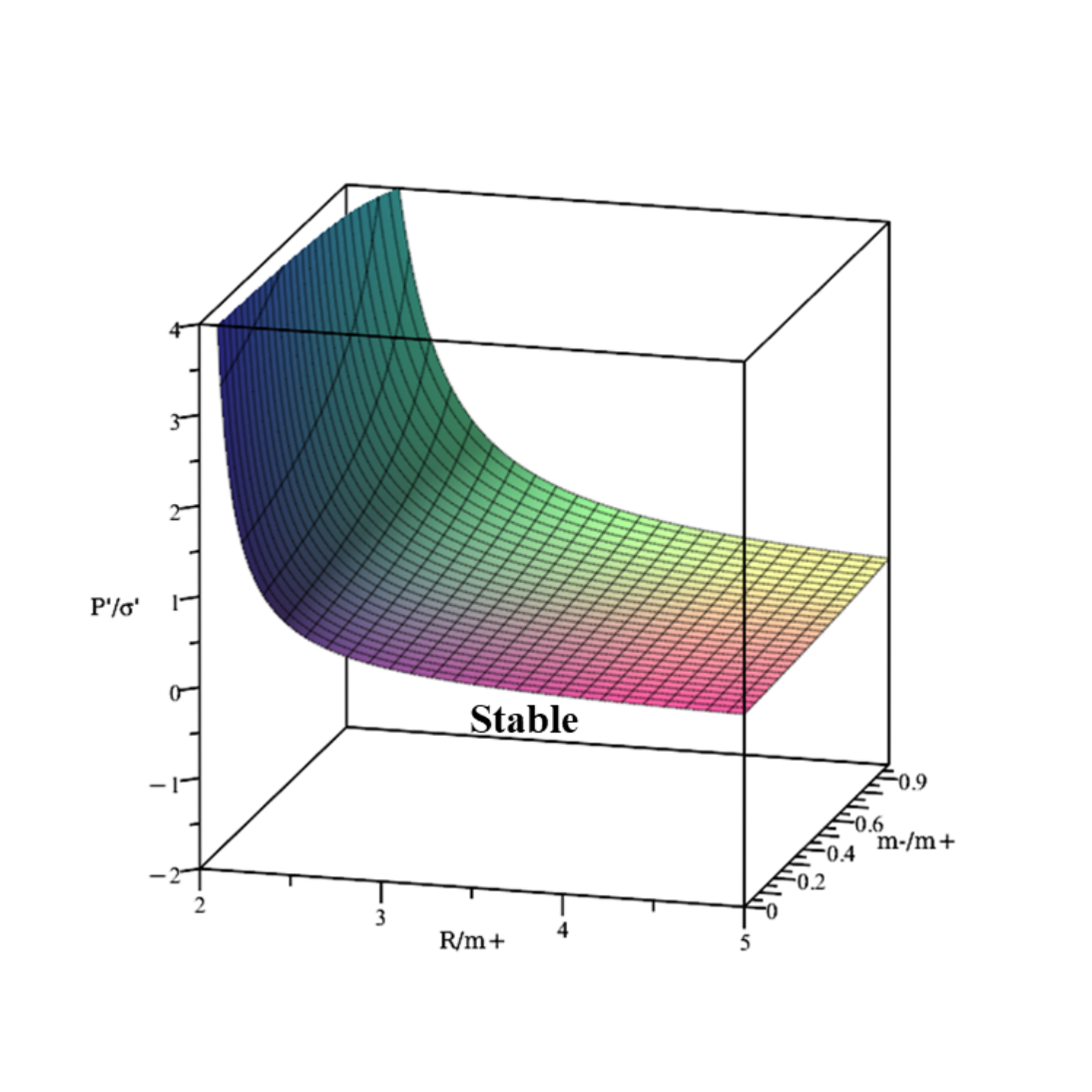}
         %\label{fig:BH}
         \caption{Black hole}
     \end{subfigure}
     \begin{subfigure}[b]{0.47\textwidth}
         \centering
        \includegraphics[width=\textwidth]{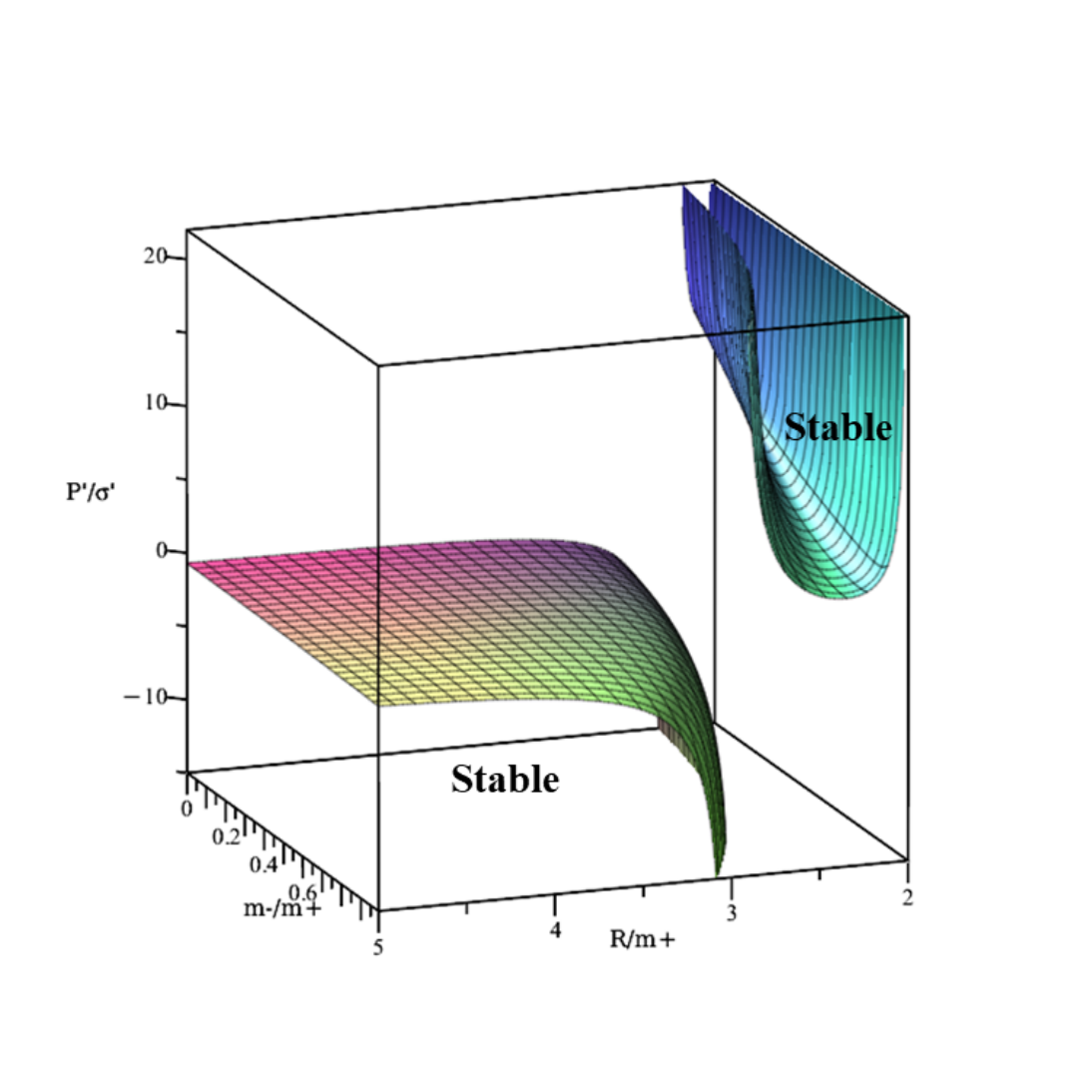}
        \caption{Wormhole}
         %\label{fig:WH}
     \end{subfigure}
     \hfill
        \caption{\textbf{Schwarzschild -- Schwarzschild Junction:} For the black hole, $M\neq0$ everywhere and there is no stability flip. Stability partially intersects the Causal Region. For the wormhole there is an asymptote present and stability is separated into two disjoint regions, neither of which intersect the Causal Region. For 2-D plot, see figure (\ref{fig:SS}).}
        \label{fig:SS_3d}
\end{figure*}
\begin{figure*}[hbt!]
     \centering
     \begin{subfigure}[b]{0.47\textwidth}
         \centering
         \includegraphics[width=\textwidth]{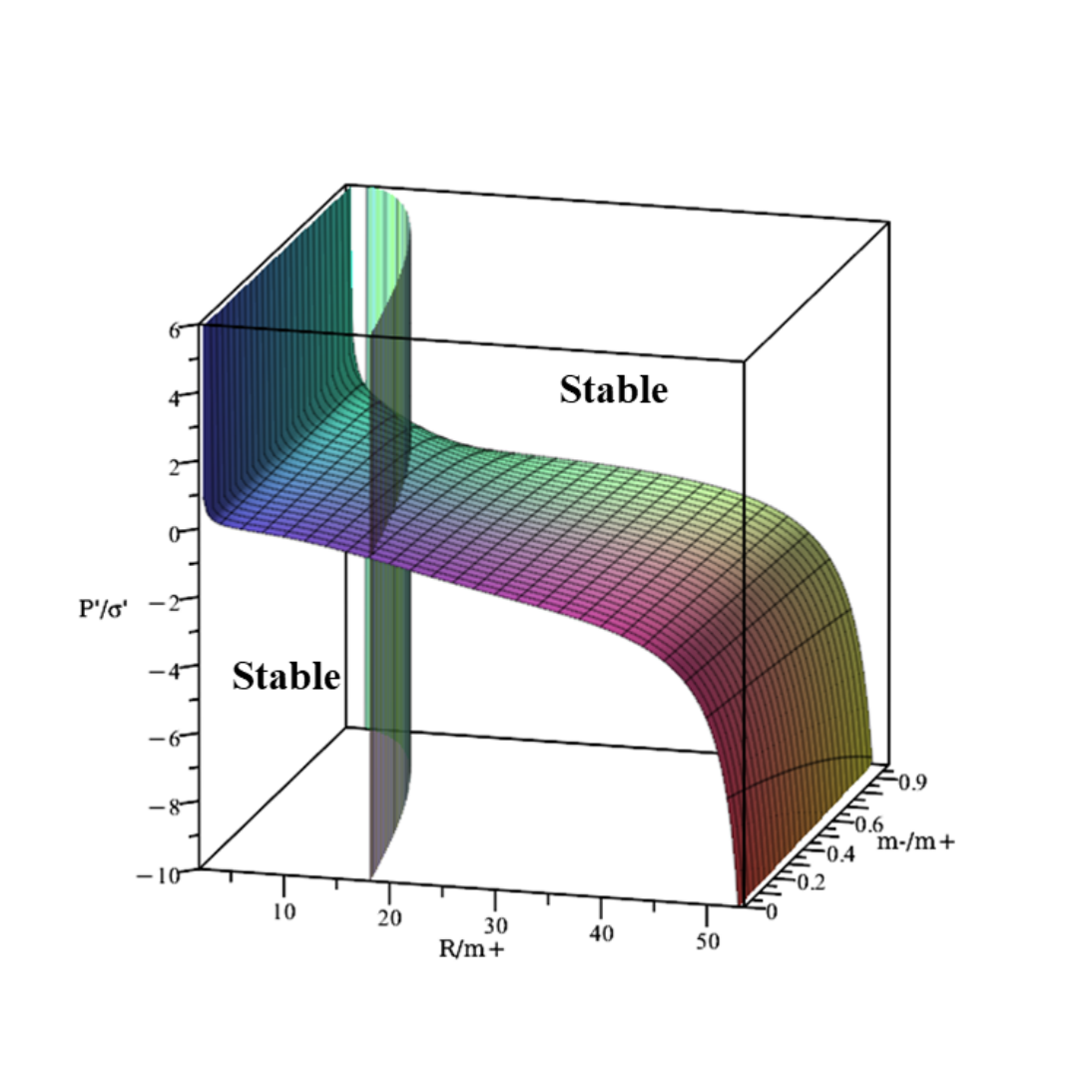}
         %\label{fig:BH}
         \caption{Black hole}
     \end{subfigure}
     \begin{subfigure}[b]{0.47\textwidth}
         \centering
         \includegraphics[width=\textwidth]{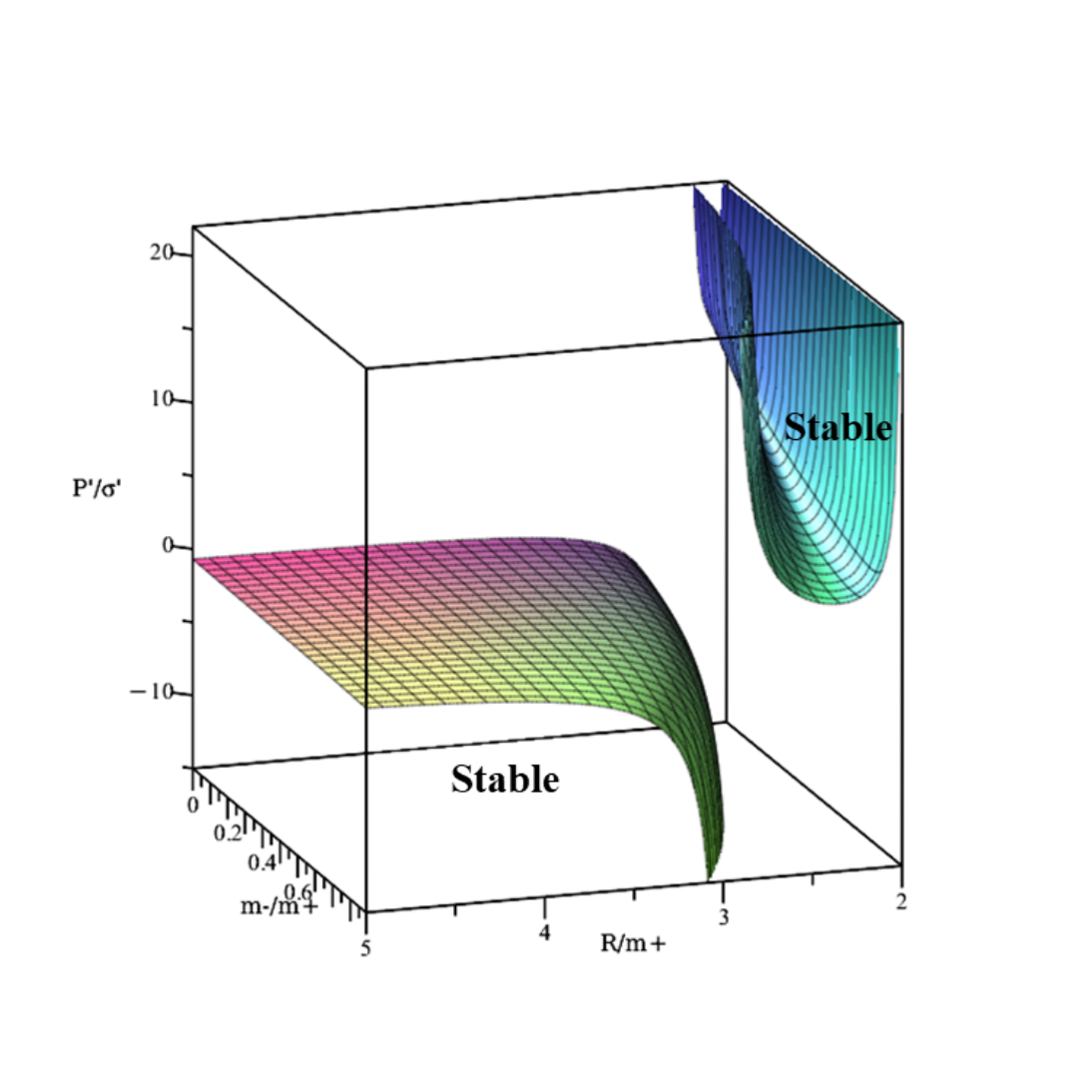}
         %\label{fig:WH}
         \caption{Wormhole}
     \end{subfigure}
     \hfill
        \caption{\textbf{Schwarzschild - de Sitter -- Schwarzschild Junction:} For the black hole, there are points where $M=0$ and a stability flip is present without an asymptote. This occurs at the vertical plane. A Schwarzschild - de Sitter -- Schwarzschild junction implies $[C] =-0.001$ which does not fulfill the asymptote condition for any beta. For the black hole plot, the de Sitter Horizon can be seen by the divergence around $\alpha \approx 54$, though the view window of the wormhole plot is too restricted for this feature to be visible. For 2-D plot, see figure (\ref{fig:SdsS}).}
        \label{fig:SdsS_3d}
\end{figure*}
\begin{figure*}[hbt!]
     \centering
     \begin{subfigure}[b]{0.47\textwidth}
         \centering
         \includegraphics[width=\textwidth]{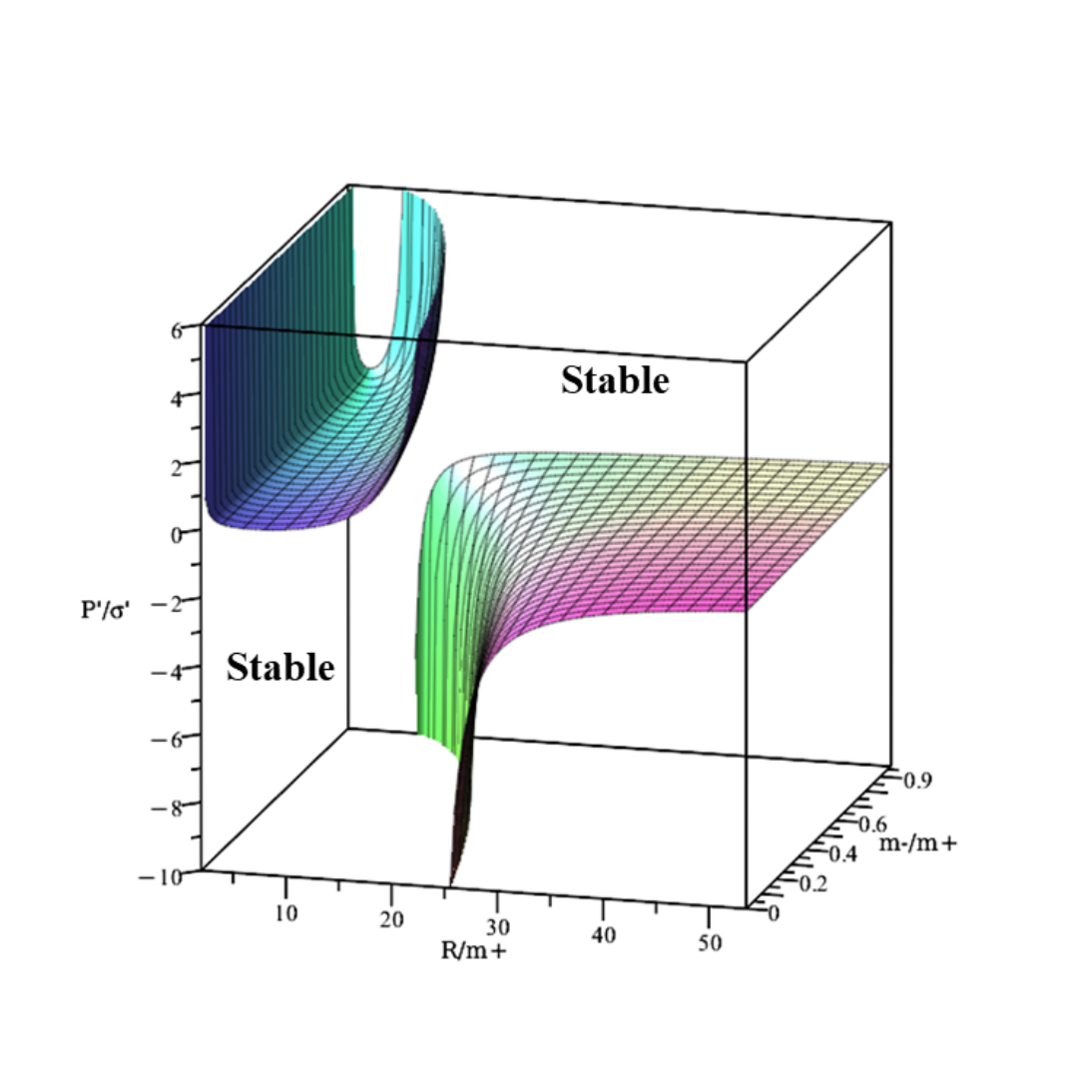}
         %\label{fig:BH}
         \caption{Black hole}
     \end{subfigure}
     \begin{subfigure}[b]{0.47\textwidth}
         \centering
         \includegraphics[width=\textwidth]{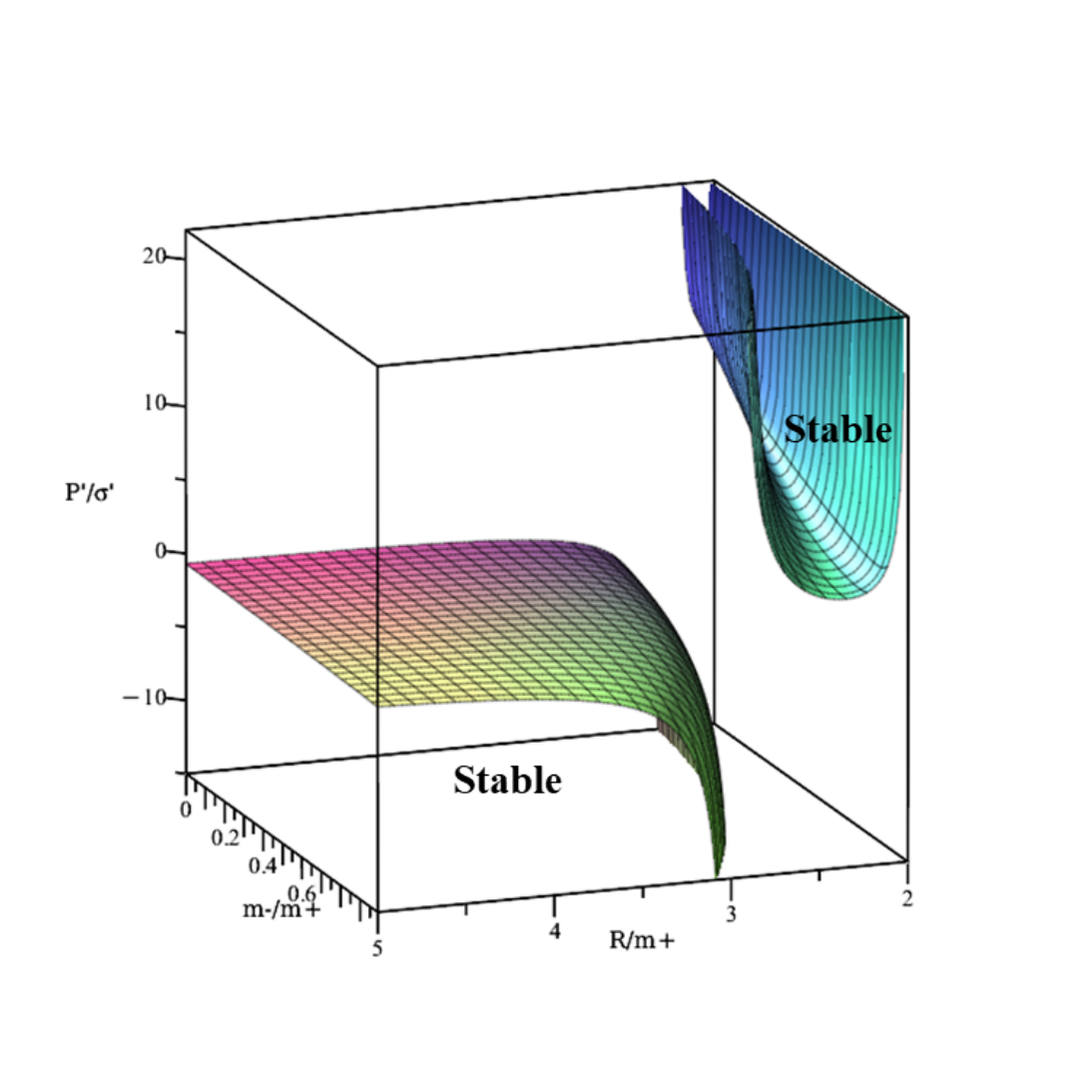}
         %\label{fig:WH}
         \caption{Wormhole}
     \end{subfigure}
     \hfill
        \caption{\textbf{Schwarzschild - anti-de Sitter -- Schwarzschild Junction:} Black hole has an asymptote as $[C] =0.001$ which does fulfill the asymptote condition. A de Sitter horizon is not present as $\Lambda^\pm\leq0$. For 2-D plot, see figure (\ref{fig:SadsS}).}
        \label{fig:SadsS_3d}
\end{figure*}
\begin{figure*}[hbt!]
     \centering
     \begin{subfigure}[b]{0.47\textwidth}
         \centering
         \includegraphics[width=\textwidth]{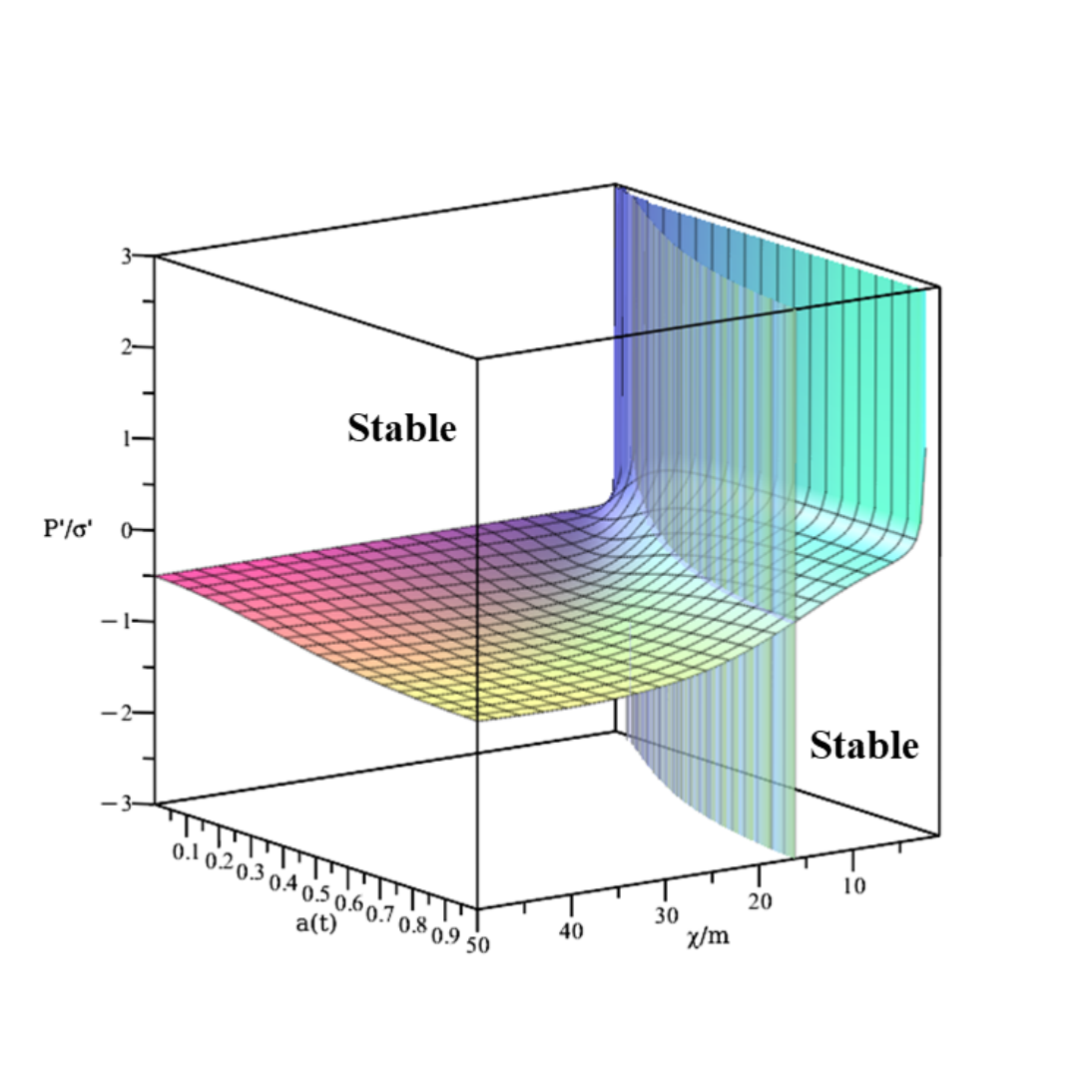}
         %\label{fig:BH}
         \caption{Black hole}
     \end{subfigure}
     \begin{subfigure}[b]{0.47\textwidth}
         \centering
         \includegraphics[width=\textwidth]{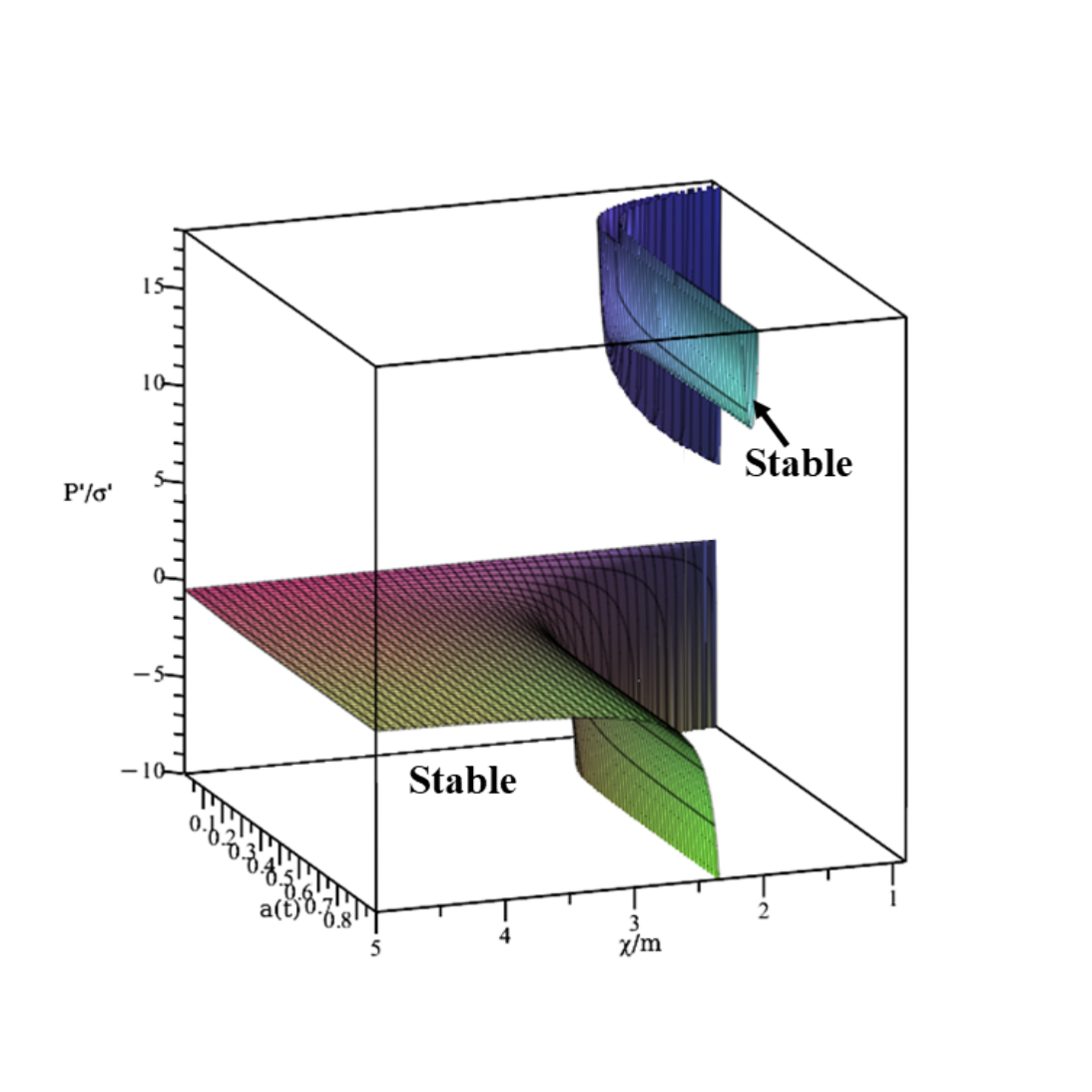}
         %\label{fig:WH}
         \caption{Wormhole}
     \end{subfigure}
     \hfill
        \caption{\textbf{FLRW (k=0) -- Schwarzschild Junction:} The black hole and wormhole graphs are similar to previous examples though does not posses a de Sitter Horizon, despite there existing a $\Lambda > 0$. It is worth noting that the radius of the event horizon decreases below $\alpha = 2$ at low $a$ due to a high value of $H^2$. For 2-D plot, see figure (\ref{fig:FS}).}
        \label{fig:SF0_3d}
\end{figure*}
\begin{figure*}[hbt!]
     \centering
     \begin{subfigure}[b]{0.47\textwidth}
         \centering
         \includegraphics[width=\textwidth]{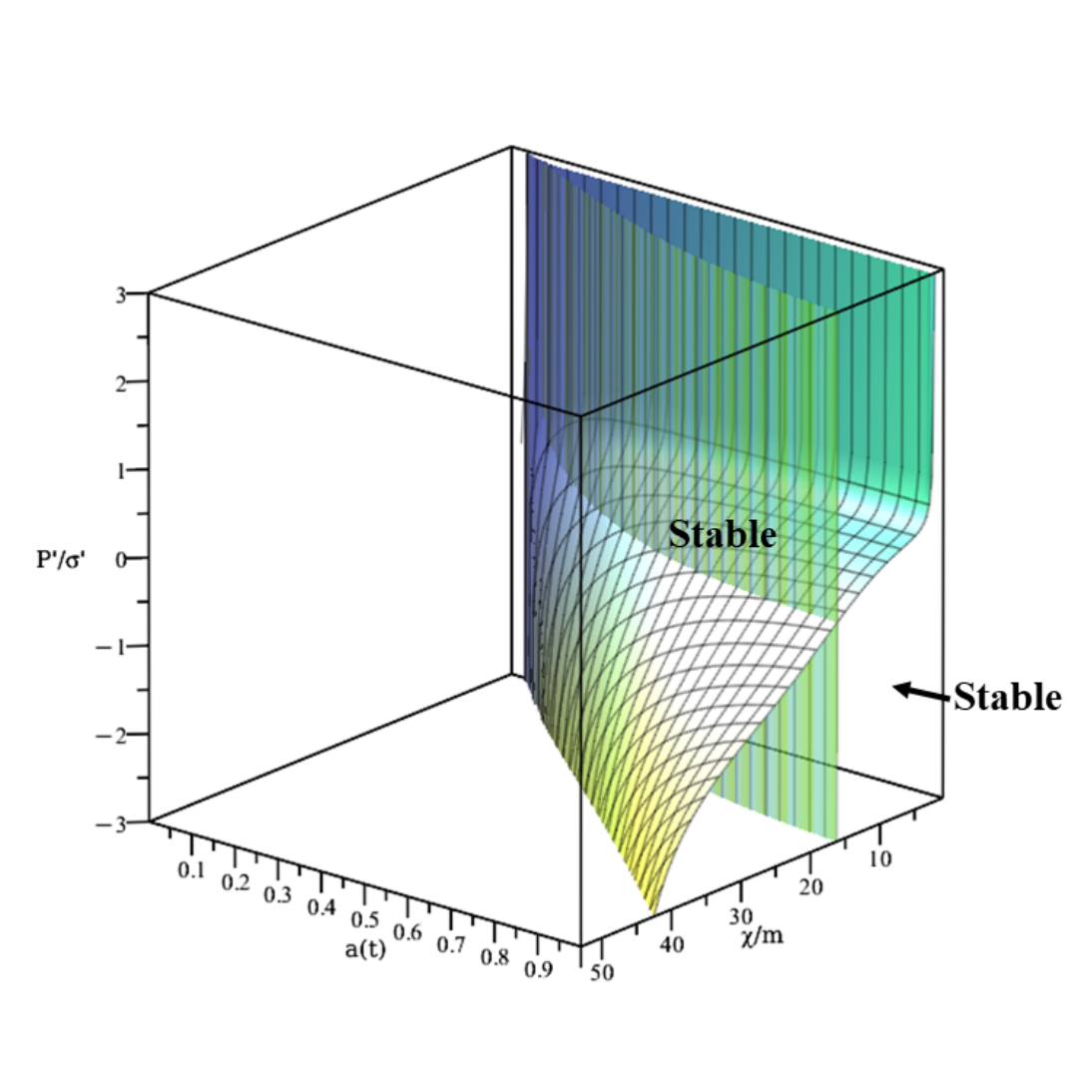}
         %\label{fig:BH}
         \caption{Black hole}
     \end{subfigure}
     \begin{subfigure}[b]{0.47\textwidth}
         \centering
         \includegraphics[width=\textwidth]{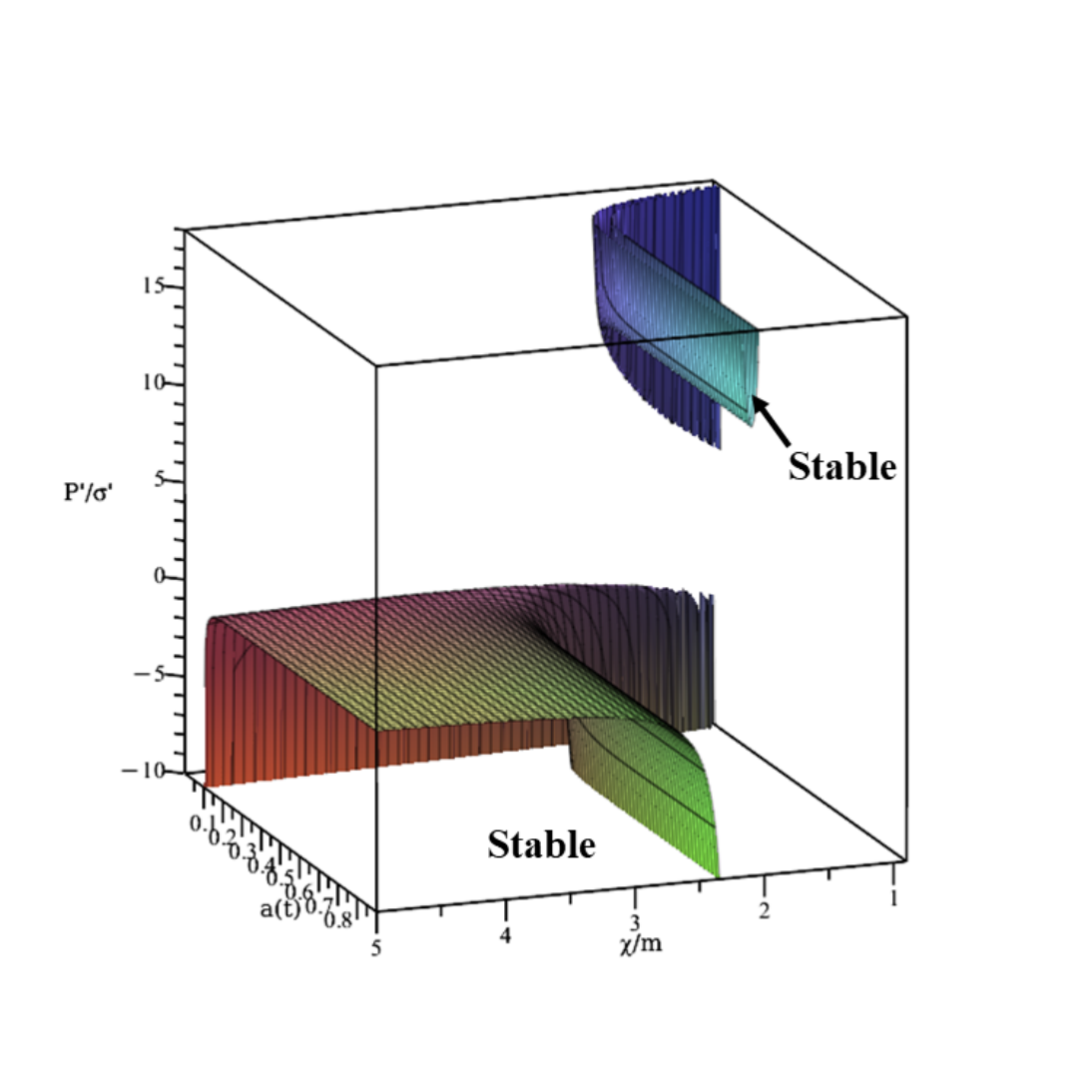}
         %\label{fig:WH}
         \caption{Wormhole}
     \end{subfigure}
     \hfill
        \caption{\textbf{FLRW (k=+1) -- Schwarzschild Junction:} For $k\neq0$ we choose $m=0.02$. It is important to note that $k>0$ gives the non physical result of $H^2<0$ for $0.420 \lesssim a(t)\lesssim 0.823$. A horizon caused by the positive curvature exists at high $\chi$ for both black hole and wormhole. At larger $m$ values the asymptote condition can be fulfilled but commonly this occurs only when $H^2<0$. For 3-D plot, see figure (\ref{fig:F+1S}).}
        \label{fig:SF+1_3d}
\end{figure*}
\begin{figure*}[hbt!]
     \centering
     \begin{subfigure}[b]{0.47\textwidth}
         \centering
         \includegraphics[width=\textwidth]{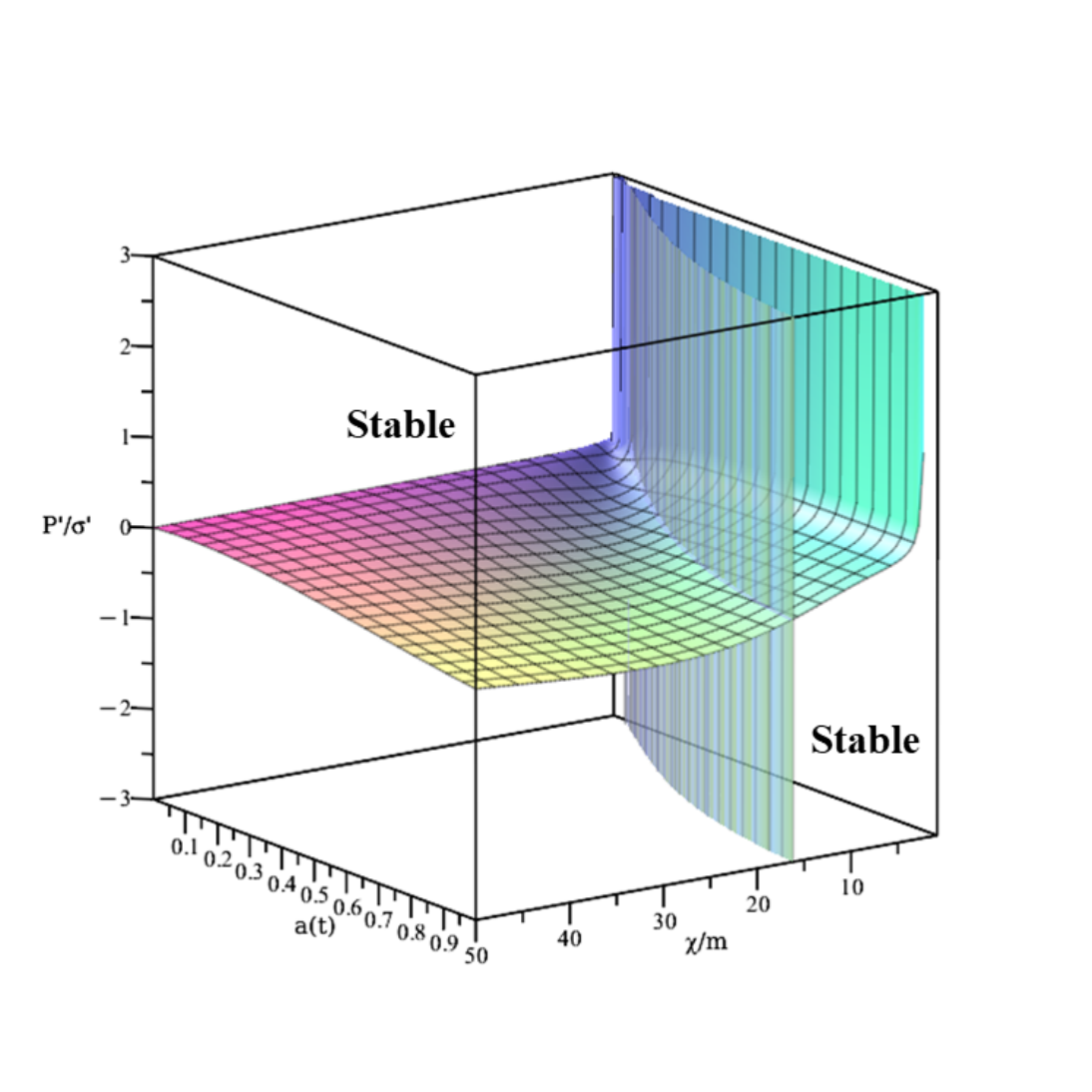}
         %\label{fig:BH}
         \caption{Black hole}
     \end{subfigure}
     \begin{subfigure}[b]{0.47\textwidth}
         \centering
         \includegraphics[width=\textwidth]{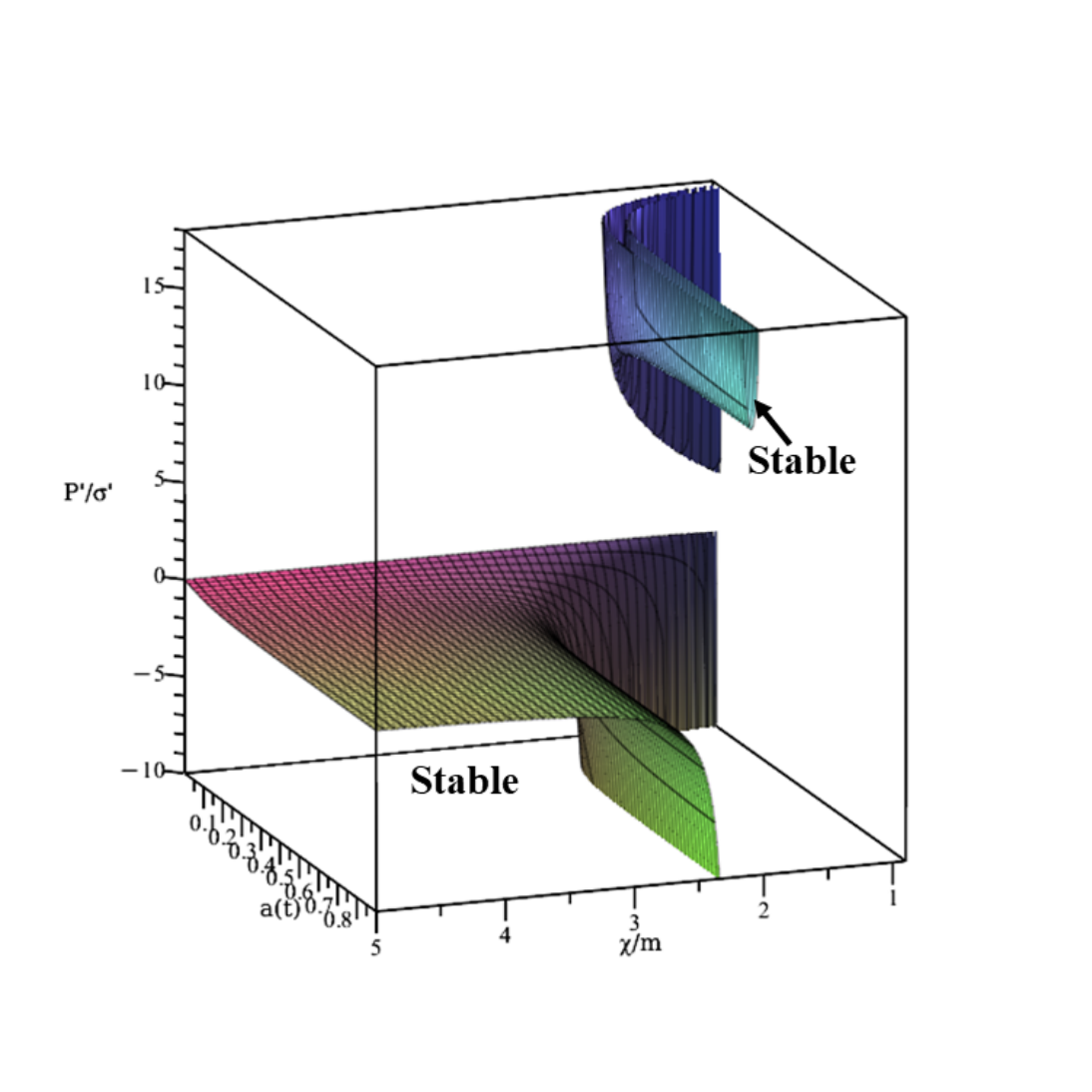}
         %\label{fig:WH}
         \caption{Wormhole}
     \end{subfigure}
     \hfill
        \caption{\textbf{FLRW (k=-1) -- Schwarzschild Junction:} There is no horizon present at high $\chi$ for either the black hole or the wormhole. As in figure (\ref{fig:SadsSads}), the stability region of the wormhole approaches the lower bound of the Causal Region as $\alpha \rightarrow \infty$. For 2-D plot, see figure (\ref{fig:F-1S}).}
        \label{fig:SF-1_3d}
\end{figure*}

In this section, the parameter spaces of a few junctions are plotted in three dimensions for a more complete visualization. For the junctions with Schwarzschild - (anti) de Sitter (figures \ref{fig:SS_3d}-\ref{fig:SadsS_3d}) $P'/\sigma'$ is plotted against the $\alpha$ and $\beta$ axes and for junctions with FLRW (figures \ref{fig:SF0_3d}-\ref{fig:SF-1_3d}), it is plotted against the $\alpha$ and $a(t)$ axes.

% \end{onecolumn}

\end{document}